%% file: main-subm-jan24.tex
\pdfoutput=1

\documentclass[alpha-refs]{nbdt-article}

\usepackage{siunitx}
\usepackage[utf8]{inputenc}
\usepackage{enumitem}

\papertype{Original Article}

\author[1,2]{Benjamin Peters}
\author[3,4,5,6]{James J. DiCarlo}
\author[7]{Todd Gureckis}
\author[8,9]{Ralf Haefner}
\author[10]{Leyla Isik}
\author[3,5,11]{Joshua Tenenbaum}
\author[12,13,14]{Talia Konkle}
\author[15]{Thomas Naselaris}
\author[16]{Kimberly Stachenfeld}
\author[1,17]{Zenna Tavares}
\author[18,19]{Doris Tsao}
\author[20,21]{Ilker Yildirim}
\author[1,22,23,24]{Nikolaus Kriegeskorte}

\affil[1]{Zuckerman Mind Brain Behavior Institute, Columbia University}
\affil[2]{School of Psychology \& Neuroscience, University of Glasgow }
\affil[3]{Department of Brain and Cognitive Sciences, MIT}
\affil[4]{McGovern Institute for Brain Research, MIT}
\affil[5]{NSF Center for Brains, Minds and Machines, MIT}
\affil[6]{Quest for Intelligence, Schwarzman College of Computing, MIT}
\affil[7]{Department of Psychology, New York University}
\affil[8]{Brain and Cognitive Sciences, University of Rochester}
\affil[9]{Center for Visual Science, University of Rochester}
\affil[10]{Department of Cognitive Science, Johns Hopkins University}
\affil[11]{Computer Science and Artificial Intelligence Laboratory, MIT}
\affil[12]{Department of Psychology, Harvard University}
\affil[13]{Center for Brain Science, Harvard University}
\affil[14]{Kempner Institute for Natural and Artificial Intelligence, Harvard University}
\affil[15]{Department of Neuroscience, University of Minnesota}
\affil[16]{DeepMind}
\affil[17]{Data Science Institute, Columbia University}
\affil[18]{Dept of Molecular \& Cell Biology, University of California Berkeley}
\affil[19]{Howard Hughes Medical Institute}
\affil[20]{Department of Psychology, Yale University}
\affil[21]{Department of Statistics and Data Science, Yale University}
\affil[22]{Department of Psychology, Columbia University}
\affil[23]{Department of Neuroscience, Columbia University}
\affil[24]{Department of Electrical Engineering, Columbia University}

\corremail{benjamin.peters@posteo.de or n.kriegeskorte@columbia.edu}

\fundinginfo{}

\runningauthor{Peters et al.}

\usepackage[most]{tcolorbox}

\newtcolorbox{abox}[3][]{
    floatplacement=#2,
    colback=black!3!white,
    colframe=black!75!white,
    enhanced,
    attach boxed title to top center={yshift=-2mm},
    title=#3,#1
}

\newtcolorbox{bbox}[3][]{
    floatplacement=#2,
    colback=black!3!white,
    colframe=black!75!white,
    breakable,
    enhanced,
    attach boxed title to top center={yshift=-2mm},
    title=#3,#1
}

\usepackage{color}
\usepackage{xcolor}
\usepackage{multirow}
\usepackage{mathtools}
\usepackage{wrapfig}
\usepackage{setspace}
\usepackage{afterpage}
\usepackage{hyperref}

\hypersetup{
    colorlinks=true,
    linkcolor=blue,
    citecolor=black,
    }

\usepackage{caption}
\usepackage{cleveref}
\crefname{figure}{figure}{figures}
\Crefname{figure}{Figure}{Figures}
\crefformat{figure}{#2Figure~#1#3}
\Crefformat{figure}{#2Figure~#1#3}

\newcommand{\itemplacement}{ht} 

\title{How does the primate brain combine generative and discriminative computations in vision?}

\begin{document}

\maketitle

\begin{abstract}
\small Vision is widely understood as an inference problem. However, two contrasting conceptions of the inference process have each been influential in research on biological vision as well as the engineering of machine vision. The first emphasizes bottom-up signal flow, describing vision as a largely feedforward, discriminative inference process that filters and transforms the visual information to remove irrelevant variation and represent behaviorally relevant information in a format suitable for downstream functions of cognition and behavioral control. In this conception, vision is driven by the sensory data, and perception is direct because the processing proceeds from the data to the latent variables of interest. The notion of ``inference'' in this conception is that of the engineering literature on neural networks, where feedforward convolutional neural networks processing images are said to perform inference. The alternative conception is that of vision as an inference process in Helmholtz's sense, where the sensory evidence is evaluated in the context of a generative model of the causal processes that give rise to it. In this conception, vision inverts a generative model through an interrogation of the sensory evidence in a process often thought to involve top-down predictions of sensory data to evaluate the likelihood of alternative hypotheses. The authors include scientists rooted in roughly equal numbers in each of the conceptions and motivated to overcome what might be a false dichotomy between them and engage the other perspective in the realm of theory and experiment. The primate brain employs an unknown algorithm that may combine the advantages of both conceptions. We explain and clarify the terminology, review the key empirical evidence, and propose an empirical research program that transcends the dichotomy and sets the stage for revealing the mysterious hybrid algorithm of primate vision.
\end{abstract}

\newpage

\tableofcontents
\addtocontents{toc}{\vspace{-.25cm}} 

\vspace{-.3cm}

\section*{Display items}
\begin{itemize}
    \item \Cref{fig:toy-example}: \hyperlink{link:fig:toy-example}{Toy example}
    \item \Cref{fig:constructing-inference-models}: \hyperlink{link-figure:constructing-inference-models}{Discriminative and generative are frameworks for constructing inference models}.
    \item \Cref{fig:examples-of-system-interpretation}: \hyperlink{link-figure:examples-of-system-interpretation}{Examples of interpreting systems in terms of discriminative and generative models}
    \item \Cref{fig:spectrum-of-hybrid-models}: \hyperlink{link-figure:spectrum-of-hybrid-models}{Spectrum of hybrid vision models}
    \item \hyperlink{link-box:models}{Box 1: The different types of `model'}
    \item \hyperlink{link-box:primer}{Box 2: A brief primer on inference with discriminative and generative models} 
    \item \hyperlink{link-box:nuances}{Box 3: Important nuances}
    \item \hyperlink{link-box:identifying-the-framework}{Box 4: Identifying the framework}
    \item \hyperlink{link-box:questions}{Box 5: Selected questions and challenges}
    
\end{itemize}

\newpage



                        


\section{Introduction}

Is vision an associative process that maps from retinal images directly to latent variables of interest? A lookup table would be unrealistic given the high dimensionality of image space, but an interpolation model employing multiple stages of representation, such as a feedforward neural network, can be very effective for visual tasks such as object recognition. In this conception, any prior knowledge about the world is implicit in the associative mapping learned for the purpose of vision. 

The alternative perspective is that vision goes beyond association, employing models that aim to capture, at some level of abstraction, the process through which the sensory data are generated. Does vision involve interpreting the sensory data in the context of an understanding of the processes in the world that generate them? This perspective suggests that vision involves inference in the context of prior knowledge. An ideal implementation would involve probabilistic (i.e., Bayesian) inference with a prior that captures all knowledge and uncertainty about the processes that produce the sensory data. 

The distinction between these two conceptions appears fundamental, and yet it is difficult to pin down to a single crisp definition if we are to avoid oversimplification. It is related to multiple important distinctions: (1) perception as inference \citep{von_helmholtz_handbuch_1867, gregory_perceptions_1980, friston_theory_2005, kersten_object_2004} versus direct perception \citep{gibson_ecological_1979}, (2) generative model versus discriminative model \citep{lecun_deep_2015, ng_discriminative_2001, lasserre_principled_2006, bishop2006pattern}, (3) mental simulation \citep{schacter2007remembering, barsalou1999perceptual} versus reverse association, (4) recurrent neural network \citep{elman_finding_1990, hochreiter_long_1997} versus feedforward neural network \citep{rosenblatt1958perceptron, lecun_backpropagation_1989, riesenhuber_hierarchical_1999}, (5) model-based control versus model-free control \citep{sutton2018reinforcement, daw2005uncertainty}. This paper aims to disentangle these often conflated dimensions and envision a research program that transcends the dichotomy.

Visual computational neuroscience aims to build neurobiologically plausible models of the information processing underlying visual perception. To provide an understanding of visual perception, such brain models must be able to perform visual tasks that primates can perform, including object recognition and scene understanding. Brain models must additionally account for detailed patterns of brain and behavioral responses across different visual stimuli and should be consistent with neurobiology, albeit abstracting from many of the details of biological implementation.

A\textit{ \textbf{brain model}} is a researcher's model of the brain, or the primate visual system, on which we focus here. Without loss of generality, we can conceptualize a brain model as having two components: an \textbf{\textit{inference model}} of the brain that describes the brain's computations in terms of representational transformations (from the retinal input to inferred descriptions of the world that contribute to successful behavior) and a \textbf{\textit{mapping model}} \citep{ivanova_beyond_2022} that relate these representations to brain activity in different brain regions or to behavioral data. A mapping model can render a brain model empirically testable based on its predictions of brain and behavioral data.  
We note that the term ``inference model'' is used here as is common in the current machine learning literature. In perception, there is assumed to be ground truth in the world. The term inference conveys that the fundamental task of a sensory system is to try to recover that ground truth. The inference process could use either a discriminative model that maps from images to latent variables or, in Helmholtz's deeper sense of the word inference \citep{von_helmholtz_handbuch_1867}, a generative model of the sensory data, which serves as the basis for evaluating the sensory evidence. 

The normative perspective provided by probabilistic inference is that the sensory evidence should be evaluated in the context of a prior over possible states of the world that we will refer to as a \textbf{\textit{world model}}. The world model could comprise a probability distribution over the sensory data (\textit{statistical world model}) and/or an understanding of causal processes in the world that give rise to the sensory data (\textit{causal world model}). However, direct discriminative mappings from images to behaviorally relevant latent variables provide attractive approximate solutions to the inference problem that sidestep the need for an explicit world model of either variety.

\renewcommand{\itemplacement}{hb}
\global\let\itemplacement\itemplacement
\input{display-items/box-models}

The space of possible inference models is high-dimensional and vast. The discriminative and generative solutions we are used to thinking about constitute diametrically opposed corner cases of the space of models. Consider two prototypical instantiations as reference points to anchor our discussion: 
\vspace{-4mm}
\begin{enumerate}
    \item A \textbf{feedforward discriminative deep artificial neural network model} that maps from images to object categories, object locations, or abstract geometrical descriptions of the scene.
    \item A \textbf{probabilistic inference model that contains a \textit{world model}}, i.e. a ``generative model'' that captures a prior over possible scenes and/or causal mechanisms in the world that give rise to the retinal data. The inference model then must implement approximate Bayesian inference on its world model, which typically requires recurrent computations.
\end{enumerate}
\vspace{-4mm}
\noindent Note that these familiar corner cases conflate multiple theoretically independent dimensions of the space of algorithms. For example, discriminative models can also be recurrent, and feedforward deep neural networks are universal function approximators and can, therefore, in principle, implement arbitrarily precise inference on a generative model \citep{van_bergen_going_2020}.

Both of the corner-case models perform what in machine learning is known as ``discriminative inference'' (technical definitions of this and other terms are given in the next Section and in \hyperlink{link-box:primer}{Box: ``A primer on inference with discriminative and generative models''}). However, the direct discriminative approach (which motivates corner case (1)) attempts to build a brain model as a mapping from sensory data to the behaviorally relevant latent variables that are the targets of inference. It attempts to directly construct the transformations needed for particular visual inferences, such as recognizing objects at the category level, identifying individuals, or representing the geometrical structure of the scene. This enables a focus on task-relevant latent variables and a restriction of the computational resources needed for inference (e.g. when a neural network of limited spatial and temporal complexity performs inference).

By contrast, the generative approach (which motivates corner case (2)) first explicitly captures the prior world knowledge that the brain brings to perceptual inference. The generative approach is motivated by Helmholtz's notion that ``objects are always imagined as being present in the field of vision as would have to be there in order to produce the same impression on the nervous mechanism'' \citep{von_helmholtz_handbuch_1867}. Note that this articulation nonchalantly injects the radical notion that perception, in fact, \textit{is} imagery: we must imagine to infer. The normative extension of this conception is the ideal of probabilistic inference, which relies on a generative model of the sensory data.  Researchers using the generative approach assume that some generative model, whether or not it is explicitly represented in the brain, underlies the inference process. The inferential computations create our percepts by combining the sensory data (likelihood) with the world knowledge (prior). Inference with a generative model promises flexibility and robust generalization.

We can thus understand each approach as an attempt to restrict the vast space of possible inference models by assuming particular constraints and priorities. The discriminative approach assumes particular inferential goals and limited computational means (such as the finite computational graph of a feedforward or time-limited recurrent neural network), flouting the principles of probabilistic inference from the get-go. The generative approach prioritizes the goal of making the best possible use of limited data, which motivates pursuing the normative and general method of probabilistic inference while treating computational efficiency as an afterthought to be addressed by suitable approximations. The computational magic of primate vision happens, we believe, in the space between these corner cases.

\section{Terminology and Scope}

A central assumption of our debate is that we can understand brain computation in terms of latent variable models. The structure of latent variable models of visual inference is as follows: When presented with an observation $\mathbf{o}$ (i.e., the retinal image or image sequence), the visual system computes an estimate over a set of latent variables $\mathbf{z}$ (e.g., the size, distance, category of an object in the visual field). These estimates can be point estimates or probabilistic estimates (see \Cref{fig:toy-example}). Latent variables $\mathbf{z}$ are encoded in neural responses $\mathbf{r}$ (for example, in the firing rates of IT neurons), and the relation between latent variables and neural responses is expressed as a mapping model (see \hyperlink{box:models}{Box: The different types of `model'} and examples in \Cref{fig:examples-of-system-interpretation}).

The inferred estimates over $\mathbf{z}$ can be the input for further downstream computations, yielding actions $\mathbf{a}$. For example, $\mathbf{a}$ may be the orientation response of an agent faced with a particular input $\mathbf{o}$ or the object category response of an observer seeing an image in an experiment. Carving up the mind into different cognitive capacities such as visual inference and decision-making is a useful simplification, even if a full understanding of why the brain cares about certain latents $\mathbf{z}$ but not others will ultimately require closing the perception-action cycle. This paper focuses on the perceptual computations: \textit{how} the brain infers latent variables. \textit{Why} the brain cares about certain latent variables $\mathbf{z}$ is beyond the scope of this paper and will require consideration of the interaction of the organism with its environment and multiple time-scales of adaptation, including learning and evolution. 

The different perspectives on the discriminative-to-generative spectrum agree that the brain computes estimates of latent variables $\mathbf{z}$. The controversy is about \textit{how} the brain computes an estimate of $\mathbf{z}$ during inference and whether we can understand these inference processes in terms of discriminative, generative or hybrid latent variable models. All models considered here perform what is known in machine learning as discriminative inference; that is they compute an estimate of the latents $\mathbf{z}$ given an observation $\mathbf{o}$, be it a point estimate of a posterior $p(\mathbf{z}|\mathbf{o})$. As a shorthand, we will refer to discriminative inference of any kind simply as \textit{inference}. 

\input{display-items/fig-toy-example}

\input{display-items/box-primer}

\clearpage

This controversy focuses on visual inference. The visual system is considered by many to have some generative capabilities (e.g., mental simulation, imagery). The question is to what extent such generative components contribute to the inference process.

\subsection{Two frameworks to construct inference models}

\renewcommand{\itemplacement}{ht}
\global\let\itemplacement\itemplacement
\input{display-items/fig-constructing-inference-models}

``Discriminative'' and ``generative'' denote two different frameworks within which one can construct inference models (\Cref{fig:constructing-inference-models}). Models can be constructed with a ``discriminative'' goal/objective to represent $p(\mathbf{z}|\mathbf{o})$, with a ``generative'' objective to represent $p(\mathbf{o}, \mathbf{z})$, or with a ``hybrid'' combination of the two \citep{lasserre_principled_2006, chapelle_semi-supervised_2006}. Generative models contain strictly more information about the relationship between $\mathbf{o}$ and $\mathbf{z}$ compared to hybrid models (see \hyperlink{link-box:primer}{Box: ``A primer on inference with discriminative and generative models''}). The frameworks differ by what they define as the \textit{constructing} goal of the model components (i.e., the objective). 

The discriminative goal (representing $p(\mathbf{z}|\mathbf{o})$) aligns with the inference goal (obtaining an estimate of $\mathbf{z}$ for a particular $\mathbf{o}$). For example, discriminative training optimizes the weights of a regression model to represent $p(\mathbf{z}|\mathbf{o})$. We can readily use these weights to perform inference, i.e. map observations $\mathbf{o}$ to latent variables $\mathbf{z}$. However, models constructed with a generative goal (representing the joint distribution $p(\mathbf{o}, \mathbf{z})$) do not align with the inference goal in a trivial way. A generative model, therefore, often needs additional computations to become an inference model. These additional computations ``invert the generative model'' to obtain $p(\mathbf{z}|\mathbf{o})$ from the representation of the joint distribution entailed in the generative model.

The computations within a constructed inference model can be described by a set of representations (e.g., a function computing the likelihood for a particular latent variable $\mathbf{z}$) and algorithms (e.g., belief propagation) (see middle level in \Cref{fig:constructing-inference-models}). The choice of the framework partly influences a modeler's choice of representations and algorithms to implement an inference model (middle level \Cref{fig:constructing-inference-models}). For example, representations of a prior and a likelihood and some iterative inversion algorithm are typical components of inference models constructed under the generative framework. However, there is substantial algorithmic and representational overlap between the two frameworks. Iterative inference can also be encountered in an inference model constructed under a discriminative framework (e.g., loopy belief propagation in conditional random field models).

\subsection{Understanding the use of the terms ``generative'' and ``discriminative''}
\label{subsec:understanding}

How should we define the terms generative inference model and discriminative inference model? We realized that referring to the (statistical) textbook definition of the use of generative and discriminative does not sufficiently capture what different researchers associate with these terms. Researchers have surprisingly diverse associations with the labels ``generative'' and ``discriminative'', mainly depending on which level (computational goal/informational or representational/algorithmic) they focus on (see \Cref{fig:constructing-inference-models}). Because models can be described at both levels, we might come to different conclusions about whether a model is generative or discriminative at the informational or representational/algorithmic level. We consider these levels a helpful device to understand the source of conceptual confusion rather than a veridical system to label particular models. 

Labeling models as ``discriminative'' or ``generative'' with the constructing framework in mind is a statement about the computational goal of the model and its contained information. When provided with the computational goal of the model, as specified by the modeler, the labeling is often straightforward. This most closely resembles the statistical definition of a generative or discriminative model. A model may, for instance, be hand-constructed to capture the joint distribution of observations and latents, and we label it as a generative model. For more complex data distributions, the construction process often involves a combination of model architecture, learning rule, training data, and optimization objective all tuned to implement the intended computational goal of capturing, e.g., the joint distribution. The notion of ``captured information'' reflects the fact that the computational goal may or may not be a good descriptor for what is actually encoded in the model after the construction process. For example, the model may (approximately) capture the information of the joint distribution (i.e., a generative model at the informational level). However, we may also imagine a model constructed with a discriminative computational goal (e.g., a recurrent ANN for object classification, trained with supervision) to capture more than just the discriminative distribution but to show emergent capture of aspects of the joint distribution. Distinguishing between the computational goal that the modeler intended for the model and the information it actually contains gives us the tools to characterize models constructed through optimization. For the remainder of the paper, we will largely consider the case where the computational goal and contained information coincide.

When presented with a computational specification of an inference model without knowledge about the computational goal or contained information, a researcher could decide to label models by their components (i.e., representations) and by the principles with which these components interact (i.e., algorithms). Let us consider two examples of inference models (mapping $\mathbf{o}$ to $\mathbf{z}$), with their computational specification. Model A contains the weights of a linear regression model that takes observations as input and returns an estimate of the latent variables $f_{\mathbf{z}|\mathbf{o}}: \mathbf{o} \rightarrow \mathbf{z}$. These weights can be understood as encoding the information $q(\mathbf{z}|\mathbf{o})$ (where $q$ refers to the Delta distribution centered on the point-estimate of $\mathbf{z}$). From the existence of such a component in model A, one might label the model as ``discriminative''. 
Model B is composed of a component $f_\mathbf{z}: \mathbf{z} \rightarrow q(\mathbf{z})$, that returns the density of $\mathbf{z}$ under the prior and another component $f_{\mathbf{o},\mathbf{z}}: (\mathbf{o}, \mathbf{z}) \rightarrow q(\mathbf{o}|\mathbf{z})$ that takes in an observation and a latent variable estimate and returns the likelihood. The model also comes with a description of the algorithm of how to compute $\mathbf{z}$ from $\mathbf{o}$ (e.g., sample multiple $\mathbf{z}^{(k)}$ and return $\mathbf{z}^*$, which has the maximum likelihood). Model B factorizes into prior and likelihood, a clear-cut example of a \textit{generative factorization}, i.e., a generative model at the representational level. 

Models constructed under the generative and the discriminative framework tend to fall into different clusters in representational and algorithmic space (middle level \Cref{fig:constructing-inference-models}). For example, discriminative models tend to be feedforward, and inference with generative models often involves iterative algorithms (e.g., Markov chain Monte Carlo). However, models constructed under the discriminative and generative framework can be feedforward or recurrent; both can use sampling-based approaches, and both can involve top-down or bottom-up processing. At the level of algorithmic and representational motifs, there is a substantial overlap between frameworks (\Cref{fig:constructing-inference-models}). This warrants caution when trying to identify whether a system is discriminative or generative from particular algorithmic or representational motifs (\hyperlink{link-box:identifying-the-framework}{Box: ``Identifying the framework''}). Finding evidence for a single algorithmic motif, like sampling, in a system may not be informative about whether this system is discriminative or generative.

\renewcommand{\itemplacement}{ht}
\input{display-items/fig-examples-of-system-interpretation}

Depending on the researchers' preferred levels of analysis from which they view a model, they can arrive at different opinions about its ``generative'' or ``discriminative'' label and there is much gray area in the middle. As an illustration for these gray areas, let's consider two other models, C and D. Model C is identical to Model A (i.e., identical weights), but we are informed that the weights enable computing the posterior under a generative model (either exactly in the case of a linear generative model or as an approximation otherwise\footnote{Note that the encoder of a VAE would be another example. The encoder is discriminative model, mapping inputs to latent variable estimates. However, these latent variables estimates represent the approximate posterior under the generative model of the VAE.}). Model D is the same as model C but, in addition, comes with another component that computes the density $p(\mathbf{o})$ for each observation under the (true) generative model (this component is not used during inference). Models B and D are clearly generative at the level of the computational goal. But only Model B factorizes into likelihood and prior and is, therefore, clearly generative at the algorithmic and representational level. Model D, like model B, is ``generative'' in that it represents the joint distribution of $\mathbf{o}$ and $\mathbf{z}$, however with a \textit{discriminative factorization} ($p(\mathbf{o})$ and $p(\mathbf{z}|\mathbf{o})$). Model A seems clearly discriminative at first sight, but it produces the same output as model C, which was constructed with a generative goal. Does the difference in intention during model construction matter for the model labels even though they are identical in their input-output behavior and their representational/algorithmic description?

\subsection{Understanding the brain in terms of discriminative and generative frameworks}

Inference models are hypotheses about how the primate visual system implements inference, which we can test by relating their internal representations (e.g., intermediate latent variables) to neural observations and behavior (see \Cref{fig:examples-of-system-interpretation}). But what do we gain by looking at the visual system through the lens of two statistical frameworks? Researchers may hold different beliefs about the scientific productiveness of this particular modeling framework. On one end, one might be convinced that the visual system literally implements a particular inference model derived from either a discriminative, generative, or hybrid framework. The success of this research program then hinges on the ability to identify the inference model (\hyperlink{link-box:identifying-the-framework}{Box: ``Identifying the framework''}). On the other end, one might believe that inference in the visual system is performed in a way that is far from any inference model derived from any of the frameworks discussed here and that we will be, at best, distracted or misled by the quest of whether primate visual inference is more discriminative or generative. A constructive intermediate position is that the discriminative and generative frameworks capture something fundamental about learning and inference that is true for any system. Interpreting primate visual inference in terms of discriminative and generative models is, therefore, a productive endeavor that helps us construct useful abstractions of brain computations, provides a framework for generalizing statistical insights to primate vision, and stimulates the construction of large sets of inference models that serve as computational hypotheses of primate vision.

\subsection{Discriminative perspective of vision}

The visual world is made up of rich, highly structured information. The primate brain can exploit these statistical regularities to rapidly extract visual signatures for even complex latent variables, such as 3D shape, physical properties of the world, and even semantic properties of an object (e.g., animacy), directly from a 2D retinal image. The discriminative approach is motivated by the idea that \textbf{the computational goal} of the visual system is to directly and rapidly extract all survival-relevant latent dimensions $\mathbf{z}$ (e.g., animacy or object category) from time-varying 2D retinal image $\mathbf{o}$ (i.e., sensory input). This approach chooses to explicitly focus its computational effort on these survival-relevant dimensions. In particular, discriminative models do not explicitly compute how often certain latent states occur in the world (i.e., the prior $p(\mathbf{z})$), how often combinations of latent states and sensory observations co-occur (i.e., the joint $p(\mathbf{z},\mathbf{o})$), or how possible observations would look for a given latent scene description (i.e., the likelihood $p(\mathbf{o}|\mathbf{z})$). This suggests that discriminative models have fewer computational requirements and constraints than models that also compute this additional information.

The discriminative approach does not specify what the survival-relevant latent variables are (i.e., what the dimensions of $\mathbf{z}$ should be), and can, in principle, be applied to any visual task. The task is typically acquired by supervised learning, where the researcher specifies the survival-relevant variables by providing a labeled set of training examples. The original conceptual models of early visual processing of Hubel and Wiesel are discriminative, where the assumed set of $\mathbf{z}$ corresponds to edges in a retinal coordinate frame. Another example of the discriminative approach is in the area of motion perception, where the assumed set of $\mathbf{z}$ corresponds to motion direction \citep{shadlen_neural_2001, britten_analysis_1992, newsome_neuronal_1989, movshon_visual_1996}.  Perhaps most notably, there is a decades-long history of discriminative work in the area of object recognition, where the assumed set of $\mathbf{z}$ corresponds to object categories or object identities \citep{fukushima_neocognitron_1980, yamins_performance-optimized_2014, yamins_using_2016, riesenhuber_hierarchical_1999}. 

It was quickly appreciated that many survival-relevant latent variables of interest in the world cannot be linearly computed from combinations of retinal inputs (i.e., from the set of photoreceptor activations). Even discriminating between just two objects requires tolerance to object position, scale, and pose, while maintaining shape selectivity. These tolerances cannot be achieved with linear functions of the input images. The discriminative approach is \textbf{algorithmically motivated} by the idea that, because primates can behaviorally discriminate among such latents, the visual system must be computing non-linear combinations of retinal inputs into more useful representations in downstream brain areas. Further, these transformations must occur rapidly, because primates make many behavioral discriminations at a time scale that is only tens of milliseconds longer than the visual latencies of neural responses in cortical regions empirically implicated in that re-representation. The discriminative approach thus engages in a search for a set of non-linear functions that can rapidly and accurately estimate survival-relevant latent variables. 

This search is usually over a family of stacked (hierarchical) linear-to-non-linear functions that approximate some of the hardware of the visual system (below) and are called artificial neural networks (ANNs). While typically feedforward, these hierarchical ANNs can also include lateral and recurrent connections. Once the parameters of such an architecture are fixed, a particular discriminative algorithmic specification is complete. That is, an estimate of $p(\mathbf{z}|\mathbf{o})$ can be computed for any input $\mathbf{o}$. We emphasize that such an algorithmic specification is partially motivated by what is known about the brain and partially by an assumed cognitive function. It is simultaneously a brain model (though highly abstracted) and a cognitive model. Discriminative ANNs currently provide state-of-the-art performance in computer-vision on a range of visual tasks, strong generalization abilities, as well as the best predictions of neural response patterns to held-out images throughout the primate visual system.

Under this framework, the process of extracting $p(\mathbf{z}|\mathbf{o})$ is a relatively stable representational coding scheme. Indeed, inducing representational changes through perceptual learning in the adult visual system is extremely challenging, requiring thousands and thousands of trials, and only induces changes that are tightly linked to the training task (e.g., training vernier acuity or intentional tracking in one location of the visual field does not even transfer to improved performance in another part of space, \cite{fiorentini_perceptual_1980}). A corollary of this stability is that visual processes are not likely to be (strongly) influenced by explicit verbal instructions like ``there is a tree to your right.''  While such linguistic input can clearly update an explicit world model, these cognitive factors do not enter into the process of seeing (related to the idea of cognitive impenetrability of visual processing, \cite{fodor_modularity_1983}). Put another way, inference $p(\mathbf{z}|\mathbf{o})$ does not involve an explicit model of the world.

\subsection{Generative perspective of vision} \label{sec:generative}

The generative framework is motivated by the intuition that our sensory information about the real world is ambiguous. Perception must consider not only the present observations but accumulated knowledge of the world \citep{von_helmholtz_handbuch_1867}. The visual system should dynamically update its beliefs about the current state of the world, and the uncertainty that remains given the observations. The generative framework formalizes these intuitions by positing that the brain encodes a probabilistic generative model of the visual world. The brain's model is a joint distribution $q(\mathbf{o}, \mathbf{z})$ over observable images, $\mathbf{o}$, and latents, $\mathbf{z}$. In the generative framework, the latent variables $\mathbf{z}$ are often assumed to represent objects or processes in the world that cause the observable inputs.

The brain's internal model of the world determines the relationships among latent variables $\mathbf{z}$ and observations $\mathbf{o}$ that define the joint distribution $p(\mathbf{o}, \mathbf{z})$. The generative framework assumes that this joint distribution is factorized in the brain as $q(\mathbf{o}, \mathbf{z})=q(\mathbf{o}|\mathbf{z}) \cdot q(\mathbf{z})$ (and not vice versa as $q(\mathbf{o}, \mathbf{z})=q(\mathbf{z}|\mathbf{o}) \cdot q(\mathbf{o})$, i.e. the discriminative direction), and that neural activity in the visual system (or some subset of it) represents the posterior distribution $q(\mathbf{z}|\mathbf{o})\propto q(\mathbf{o}|\mathbf{z}) \cdot q(\mathbf{z})$. The commitment to this particular factorization assumes that changes that occur slowly and are learned (e.g., synaptic or structural changes) reflect changes to the parameters of $p(o|z)$ and $p(z)$, while changes in activity reflect updates to $p(z|o)$, which varies as quickly as $o$ varies.

There are multiple normative advantages to learning the generative rather than the discriminative factorization of the joint distribution. One of the advantages is that learning $p(\mathbf{o}|\mathbf{z})$ enables the brain to approximate the causal dependencies of the external world. The observations $\mathbf{o}$ are physically caused by unobserved latent variables $\mathbf{z}$ and not vice versa. The advantages of a factorization that reflects causality, compared to a non-causal discriminative factorization $p(\mathbf{z}|\mathbf{o})\cdot p(\mathbf{o})$, are reviewed in \citep{scholkopf_statistical_2022}. They include higher generalization performance due to greater invariance to different contexts. For example, when transitioning from indoor to outdoor spaces, the kind of light source changes, but the physical relationship between light and images remains constant. Such contextual changes will manifest themselves in $p(\mathbf{z})$ rather than changes in the physical relationships approximated by $p(\mathbf{o}|\mathbf{z})$. Furthermore, $p(\mathbf{o}|\mathbf{z})$ will be more modular, i.e. its individual parts will be more independent of each other, and any changes within these parts under distributional shifts will be sparser \citep{scholkopf_statistical_2022}. These properties are especially important considering the fact that the parameters of whatever factorization the brain learns are likely stored in synapses that change more slowly than context (e.g., when moving from an indoor to an outdoor environment), and need to be learned through mostly local learning rules. Finally, causal models allow for predictions of the consequences of causal interventions (key for action selection) and for the evaluation of hypothetical statements (e.g., beneficial for offline learning; \cite{hamrick_role_2021, liu_human_2019}).

The relationship between the brain's model of the world and the actual world may be expressed through the match of the brain's marginal distribution over observations $q(\mathbf{o})=\int q(\mathbf{o}|\mathbf{z})q(\mathbf{z}){\rm d}z$ and the distribution over observations generated by the world, $p_{\rm world}(\mathbf{o})$. A necessary, but not sufficient, condition for this match is the so-called ``calibration requirement'' \citep{pakdaman_naeini_obtaining_2015}. Let $p_{w}(o,\tilde{z})$ be the joint distribution over observable images and their causes $\tilde{z}$ in the actual world. The `actual causes' $\tilde{z}$ are inaccessible and the brain instead models the world with its own latent variables $\mathbf{z}$. The brain cannot assess the viability of its $\mathbf{z}$ w.r.t. to $\tilde{z}$ directly but only through a calibration requirement: $q(\mathbf{z}) = \mathbf{E}_{o\sim p_{w}(\mathbf{o})}[q(\mathbf{z}|\mathbf{o})]$ (as well as the effect of its own actions, which we do not consider here).

The $\mathbf{o}$'s and the $\mathbf{z}$'s differ not only in their observability. The $\mathbf{z}$'s obtain the status of ``generators'' by virtue of the conditional independence assumptions specified by $q(\mathbf{o}, \mathbf{z})$. 
If the brain has an ability (e.g., cognitive) to influence its beliefs over subsets of $\mathbf{z}$, then it can use its internal model to compute the implied beliefs over other subsets of $\mathbf{z}$'s.
Many visual tasks require one to flexibly alter the representation of a subset of the $\mathbf{z}$ in a way that is consistent with varying assumptions or beliefs about other $\mathbf{z}$'s. For example, to choose a new hairstyle, one might observe one's face in a mirror while mentally superimposing different hairstyles, and judging the aesthetic attributes of each. In the generative framework, the visual system supports such tasks by making inferences that are conditioned upon both observations and \emph{hypotheses}. Formally, this means computing posteriors that include latents in the conditioning set  $q(\mathbf{z}_A | \mathbf{o}, \mathbf{z}_{\setminus A})$, where the values of the conditioned-upon latents $\mathbf{z}_{\setminus A}$  express assumptions or beliefs. In the hairstyle example, $\mathbf{z}_A$ specifies aesthetic attributes, and $\mathbf{z}_{\setminus A}$ specifies hairstyle. By conditioning on $\mathbf{z}_{\setminus A}$, the brain could construct representations, $p(\mathbf{z})$, of a hypothetical world. The generative framework thus provides a very natural (and normative) explanation for purely generative events, like mental images \citep{breedlove_generative_2020}, which are detectable when conditioned-upon beliefs vary while an observed image is held fixed (e.g., by closing one's eyes).

\subsection{Beyond the dichotomy: Hybrid models of vision}
\label{sec:terminology:hbyrid_models}

The generative and discriminative frameworks engage the common goal of discriminative inference of $p(\mathbf{z}|\mathbf{o})$. However, the generative framework additionally engages other types of inference that the primate and, in particular, the human brain are capable of. Comparing and combining the two frameworks benefits from a focus on the shared goal of discriminative inference of $p(\mathbf{z}|\mathbf{o})$. We consider the question posed in the title of the paper: How does the brain combine generative and discriminative computations? (\Cref{fig:spectrum-of-hybrid-models})

The notion that the brain combines elements of both frameworks suggests the possibility of hybrid models, which get the best of both worlds or at least strike a favorable compromise, trading off some of the statistical efficiency of the generative approach to achieve better computational efficiency as afforded by the discriminative approach. The idea that some compromises are required is not unique to brain science. Computer vision engineers as well have long realized the advantages of both approaches: the theoretical advantages of solving vision with generative models and probabilistic inference and the practical need to find efficient implementations that more directly map images to latent variables of interest as achieved by feedforward neural network models.

\textbf{Division of labor by level of representation.} One way to combine discriminative and generative components is by using them at different levels of representation \cite[see e.g.,][]{Koblinger_representations_2021}. Discriminative computations could serve to rapidly compute a summary of the sensory evidence that renders generative inference more tractable. The discriminative summary could, for example, be invariant to nuisance variables that are accidental to the viewing conditions and irrelevant to behavior, such as image contrast and specular reflections. It could also provide first guesses about semantic information, such as object category and shape, as well as relational information, while somewhat separating and perhaps de-emphasizing less essential information such as the precise position of each object and appearance details. Generative inference could then operate on the summary statistical representation, treating it as the observed data, and solving a computationally less costly inference challenge.

This idea clarifies the tradeoff between statistical and computational efficiency: Summarizing the evidence risks loss of information, but promises to reduce the cost of generative inference. Although this idea may capture an important element of how the primate brain leverages hybrid inference, it seems questionable whether the visual system can be cleanly divided into a lower-level discriminative and a higher-level generative module. The literature suggests that both discriminative and generative forms of inference may be used not only in higher stages of representation, but even in primary visual cortex. Moreover, the advantages of both generative and discriminative inference do not depend on the level of abstraction of the representations, suggesting that hybrid inference might be useful at all stages.

\textbf{Division of labor by temporal stage of processing, generalization challenge, and energetic and time constraints.} Another approach to marrying the two is by using a discriminative model (often a feedforward neural network, but not necessarily so) for fast amortized inference and refining the inferences with generative inference components when this promises improvements and the time and energy for those additional computations is available. Amortized inference can be understood as the crystallization in a memorized mapping of inference components that are frequently needed \citep[also see][]{shelton_select_2011} for an algorithm in which the relevant components are selected dynamically, depending on the observation). This provides a way for a system to continually optimize the efficiency of its own inferential computations, while retaining the ability to revert to more laborious iterative inference computations when these are needed to perform well. Beyond storing a look-up table, amortized inference can also generalize to new inputs. More laborious iterative inference can exploit the advantages of generative inference and is expected to help (a) in novel situations, where more ambitious forms of generalization are needed, (b) when inputs are rendered ambiguous through occlusion, noise, or low contrast, and (c) when the accuracy of the perceptual inferences is paramount and/or the resources needed for more computation, time and energy, are readily available.

An attractive framework for conceptualizing hybrid inference is iterative amortized inference \citep{marino_iterative_2018, emami_efficient_2021, greff_multi-object_2019, tschantz_control_2020}. In this framework, amortized inference is used not just to map from the sensory evidence to the result of discriminative inference, but from the combination of the sensory evidence and its current interpretation to an improved interpretation of the sensory evidence. In the first step, the system performs discriminative inference. Subsequent steps can refine the inference and incorporate generative elements such as predictions of the evidence on the basis of the current interpretation whose errors drive adjustments to the interpretation. A model of this type would be best understood as discriminative early after stimulus onset but would benefit from the advantages of generative inference at later stages.

\input{display-items/fig-spectrum-of-hybrid-models}

\textbf{Synthesis models.} Instead of building separate discriminative and generative inference components and getting them to work as a team, evolution and learning more likely developed a seamless synthesis of mechanisms we conventionally associate with discriminative and generative inference. Can we envision brain computation in a language that transcends the division? What if the brain learns to infer (1) the future from the past (simulation) and (2) the past from the future at multiple levels of abstract representation? What if it also learns to infer (3) more abstract from less abstract representation and (4) less abstract from more abstract ones? What if it similarly infers representations of features or objects in one location from representations of features and objects at a different location, at multiple levels of abstraction? An interactive process could achieve rapid approximate inferences that are iteratively refined to the extent that the required time and energy are available.

\newpage
\renewcommand{\itemplacement}{ht}
\input{display-items/box-imporant-nuances}

\newpage

\section{Relevant empirical observations}

\subsection{Behavioral observations}

\subsubsection{Recognition reaction times}

\paragraph{Phenomena}

\textit{Recognition is rapid.} Very briefly presented images ($\leq$ 50 ms, in some studies as little as 13 ms, \cite{potter_detecting_2014}) followed by visual masks or shown in a rapid-serial-visual-presentation (RSVP) stream suffice to extract rich semantic information \citep{thorpe_speed_1996, serre_feedforward_2007,potter_meaning_1975,greene_briefest_2009}. Classification reaction times can be short (150-200 ms \cite{thorpe_speed_1996}) allowing little time for recurrent computation.

\vspace{3mm}
\noindent\textit{Reaction times depend on the stimulus and task set.} Reaction times and recognition performances vary across images \citep{spoerer_recurrent_2020} and under different viewing conditions \citep{kar_evidence_2019, rajalingham_comparison_2015, rajalingham_large-scale_2018}. 
More difficult images take longer to recognize \citep{kar_evidence_2019}.
There is a speed-accuracy trade-off \citep{wickelgren_speed-accuracy_1977}: When faster responses are required of the subject, the accuracy with which objects are categorized suffers. It thus appears that both bottom-up (image difficulty) and top-down (cost of delayed recognition) constraints affect the reaction time in primate object recognition.

\paragraph{Discriminative Interpretation}
Accurate recognition of briefly presented and backward masked images is consistent with largely feedforward, discriminative computations. Physiological evidence suggests that masked stimuli continue to be processed in a feedforward manner along the ventral visual hierarchy, but that masking blocks reentrant or top-down processing (e.g., from IT to V1) \citep{rolls_processing_1997, fahrenfort_masking_2007}. Despite this disruption to top-down processing in RSVP paradigms, recognition performance is still well above chance, suggesting that feedforward processing is sufficient to perform most high-level visual inference.

Under challenging viewing conditions, rapid recognition performance decreases and reaction times slow. Feedforward discriminative models may not predict this slowing of responses but can be reconciled with it: If neural representations require some time to ramp up at each stage \citep{kietzmann_recurrence_2019}, even a feedforward model can account for the variability of reaction times. This admittedly requires at least minimal forms of internal state (or equivalently recurrence), but does not require abolishing the discriminative class of models. In fact, the best models to match primate recognition accuracy and reaction times, even for challenging images, are still discriminative, though often recurrent \citep{spoerer_recurrent_2020, kar_evidence_2019}. The relatively rapid timing of all these recognition phenomena is a challenge for generative proposals of vision that require explicit (and often iterative) comparison between a top-down target and input stimulus properties.

\paragraph{Generative Interpretation}

In typical hierarchical generative models, feedback signals between different layers communicate expectation signals (priors), while feedforward signals communicate likelihoods, posteriors, or error signals (depending on model flavor). Studies that report short recognition times have used high-contrast images with little ambiguity about task-relevant content. For such stimuli, the corresponding likelihoods are highly informative and dominate prior information with the consequence that they cannot distinguish between discriminative and generative models: in both cases, stimulus information is propagated from lower to higher levels with either no (discriminative models) or little (generative models) interference from feedback signals.

At the same time, recognition times predicted by feedforward models are -- by construction -- independent of image content, and in contradiction to a large body of empirical results. In contrast, inference in generative models of visual perception is usually assumed to be iterative in nature, and as such \emph{in principle} capable of explaining image-dependent recognition times. Whether iterative brain models based on generative models of visual inputs, or iterative brain models based on discriminative computations, better account for recognition times, and neural dynamics, remains an open empirical question.

\paragraph{Hybrid models}

Recent work \citep{kar_evidence_2019} has demonstrated that adding feedback and recurrent connectivity to deep neural networks can account for of the variance in reaction times across different images. It is currently unclear whether these, mechanistically defined, models have a generative or discriminative interpretation (see \hyperlink{link-box:questions}{Box 5}), or whether improving them further to explain \emph{all} of the reaction time variance will result in models that are best understood in either the discriminative or generative perspective.

\subsubsection{Visual imagery and aphantasia}
\paragraph{Phenomena} 
Most people experience some form of internally generated imagery, and at least some of these forms are known to engage the visual cortex. For example, voluntary visual imagery engages all of the visual cortex \citep{winlove_neural_2018}, and even the thalamus \citep{chen_human_1998}. Visual encoding \citep{naselaris_voxel-wise_2015} and decoding \citep{thirion_inverse_2006} models generalize well from vision to imagery, and even from vision to dreaming \citep{horikawa_neural_2013}, indicating shared representations between external and internally generated imagery. Damage to the visual system can result in involuntary forms of imagery (e.g., hallucinations, flashbacks), and there is evidence that such imagery experiences also activate the visual cortex \citep{breedlove_or_2022}.

Thus, the human visual system clearly has generative capabilities, in the sense that it can and does generate activity that encodes visual representations that are independent of retinal input. The existence of this generative capability raises the question of whether it plays an essential role in seeing. We address this question from the discriminative and generative perspectives, respectively.

\paragraph{Discriminative interpretation}

The generative capabilities of the human visual system can be accommodated by the discriminative perspective if mental imagery is viewed as an extravisual process or one of the many varieties of noise in the visual system (i.e., activity in the visual system that is independent of retinal input). Under this view, imagery is a special kind of noise that is “on manifold”, and could be generated outside the visual system. Such noise could be functional: for example, extra-visual systems could hijack the visual system by routing their process-specific variables through the purely discriminative network that the visual system houses, effectively using visual representations as part of the solution to tasks other than seeing. In this scenario, the generative capability of the visual system would be independent of seeing, and could obstruct seeing, even while offering benefit to other systems that did not evolve to process retinal input. The challenge for this perspective is that it assumes a generative/world model outside of the visual system that is capable of generating the activity of \textit{low-level}, detailed, representations that are then propagated forward to give rise to the desired abstract categories (e.g. the orientated edges necessary to make up hair).

Support for this perspective comes from the observation that voluntary mental imagery, at least, is, by definition, independent of vision. Unless the visual input is very weak \citep{Dijkstra_subjective_2023}, imagery can operate in parallel with vision without disturbing it. Further support comes from facts that (1) signal variation across acts of imagining is much weaker than signal variation across seen images, (2) the strength of the imagery signal, as well as the salience of the experience of imagery, varies across individuals much more than visual signal or visual experience. This is most evident in cases of aphantasia, where individuals report no subjective visual perception during attempts at mental imagery but have normal visual experiences \citep{zeman_lives_2015}.  Collectively, these observations suggest that imagery is more like a cognitive strategy for satisfying extravisual objectives than a signature of generative processes working for the benefit of vision.

\paragraph{Generative interpretation}
We consider a weak and a strong interpretation of the relationship between imagery and seeing. Under the weak interpretation, generativity is essential for seeing, but \emph{imagery} is not. The existence of imagery proves that the human visual system houses a model of the world, but imagery has no essential function (aside from its epistemological role as a piece of subjective evidence that motivates the generative perspective). Imagery is thus again one of the many varieties of noise, and is a small price to pay for housing the model of the world that we need in order to see.

The strong interpretation of generativity is that it is for \emph{more} than seeing. Under this interpretation, seeing is just one form of inference the visual system performs. It performs many others, i.e., it can encode (parametrically or through sampling) all possible posterior distributions, all related through the joint distribution $p(o, z_1, z_2)$. Seeing is the inference $p(z_1, z_2 | o)$, and pure mental imagery can result from the inference $p(z_1 | z_2 = \mu_2(o^*), o=0)$, where $\mu_2(o^*)=\mathbb{E}_{z_2
\sim p(z_2|o=o*)}[z_2]$. The graph of $p(o, z_1, z_2)$ determines how the representations formed in each case relate to one another. Generative capabilities are needed to perform this kind of flexible inference. 

Support for the strong generative perspective comes from the observation that, while imagery is independent of vision, many tasks require simultaneous seeing and imagining (see the “hairstyle” example above). This is consistent with a system in which the retina is just one of the possible sources of conditioning information that is incorporated into multiple forms of inference. More concrete evidence comes from the finding that although seen and imagined features may be the same, tuning to seen features is distinct from tuning to imagined features \citep{breedlove_generative_2020, favila_perception_2022}, as would be expected if vision and imagery were distinct forms of inference. Note that this strong generative perspective subsumes portions of the weak perspective, in the sense that it assumes that generative capabilities are part of seeing. In support of this, \cite{breedlove_generative_2020} reported that when seeing and imagining are treated as different forms of inference with the same generative model, the differences in tuning to seen and imagined space and features are recovered.

\subsubsection{Perception is impenetrable by cognition}
\label{sec:empirical:perception_is_impenetrable}
\paragraph{Phenomena} 
\textit{Vision is not influenced by cognition.} Visual representations have long been thought to be separate or ``encapsulated'' from other cognitive functions \citep{fodor_modularity_1983}. In other words, explicit world knowledge from non-visual domains, like verbal instructions or emotional state, have little to no effect on visual perception. This can perhaps be seen most clearly with visual illusions. Knowledge of an illusion (e.g., that two squares are the same color in a lightness illusion, or two lines are the same length in the Müller-Lyer illusion) does not eliminate the illusory percepts \citep{firestone_cognition_2016}. 

Visual representations also have many distinctive behavioral hallmarks, including rapid and automatic processing, adaptation, attentional capture, visual search, and recognition advantages, which separate them from other cognitive functions. In this way, researchers have argued that many high-level aspects of vision, including animacy \citep{gao_psychophysics_2009} and causality \citep{rolfs_visual_2013} are extracted in a purely visual manner. 

\paragraph{Discriminative interpretation}
Explicit knowledge or cognitive states shape and can change our internal world model. For example, knowledge of the physical world or an observer's emotional state can both impact explicit judgments of a visual scene (e.g., our understanding of task difficulty or trait judgments). While these factors affect our cognitive judgments, they do not change visual percepts of such scenes \citep{firestone_cognition_2016}. This separation between our internal model of the world and what we see is compatible with bottom-up, discriminative theories of vision.

\paragraph{Generative interpretation}
The visual system's internal model of the world need not be based on explicit cognitive knowledge. Instead, these causal, top-down factors could be represented implicitly and/or contained within the visual system. As such the visual system could be both generative and ``encapsulated'' from higher-level aspects of cognition.

\subsubsection{Vision generalizes despite the open-ended compositionality of the visual world}

\paragraph{Phenomena} 

\textit{Seeing objects draped under cloth:} How well does the visual system generalize at the “long-tail” of what can happen in the visual world, in terms of the perception of shapes and other object attributes, and how does this constrain computational accounts of vision? A recently studied example of this is the perception of the shape of an object that is draped and occluded under a cloth. This is remarkable, because despite the fact that none of the visible surfaces in the image belong to the object, we can perceive many aspects of the object's shape, in addition to its category, pose, and size. Critically, no top-down or explicit shape-related tasks are necessary for the visual system to see objects under cloth – instead, this seems to be a process that is reflexive and automatic, similar to seeing the color of a surface. 

\noindent \textit{Seeing in the dark:} A similar phenomenon occurs in images with very little “evidence” – images in which most of the pixels are simply turned off or have low luminance (due to dim lighting). A prominent example is the two-tone, thresholded black-and-white images, also known as the Mooney images \citep{mooney_age_1957}. Human observers experience vivid percepts of three-dimensional shapes in such images, despite the lack of contour, surface, and texture information \citep{moore_recovery_1998}.

\paragraph{Discriminative interpretation}
One possibility is that the visual system might engage a large set of discriminatively trained features to generalize broadly. Recent computer vision models based on deep convolutional neural networks (DCNNs), trained to discriminate object category labels, provide learned feature hierarchies that can yield impressive object recognition capabilities. These feature hierarchies are relatively robust to certain kinds of variation, including pose and background, even though the training objective does not explicitly include these goals. Moreover, these same features can be used for many downstream visual tasks with only minor adaptation (e.g., fine-tuning) \citep{zhang_unreasonable_2018}. It is possible that these feature hierarchies are also sufficient to generalize across even more extreme image transformations, such as cloth occlusion or Mooney images.

\paragraph{Generative interpretation}

A different possibility is that we see 3D shape via “analysis by synthesis”, or inference in a physics-based generative model of how scenes form and give rise to images \citep{yuille_vision_2006, erdogan_visual_2017, mumford_neuronal_1994, heyer_pattern_2002}. 
According to this perspective, shape perception is not solely or primarily determined by a fixed, universal collection of image cues that are computed in a bottom-up manner from any image and are adequate for any downstream task. Instead, 3D shape percepts are inferred through a top-down synthesis process recruiting an internal model of how physical scenes form and project to images. The generative model approach considers cloth draping as only one instance of an unbounded range of uncommon object presentations where certain characteristics of the physical properties of scenes and images can significantly modify an object's appearance from its typical form, yet remain easily understandable by humans. Similar examples include viewing objects such as chairs or airplanes outside in a rainstorm, under ten feet of cloudy water, behind colored plastic wrap, or in the light of a full moon at night. This open-ended compositionality of the visual world suggests that a system should model individual, scene-level causes and how they combine and project to sensory inputs. Then, reversing the effects of these causes may recover the original physical scene. Thus, the visual system could still identify a draped object and some of its fine-grained 3D shape by modeling and inverting “cloth physics”. 
Similarly, the visual system could perceive 3D shapes in Mooney images by modeling extreme illumination conditions that could yield two-tone, black-and-white images (or simply a thresholding filter) and by inverting this generative process during inference. 

\paragraph{Hybrid approaches}

In addition to purely discriminative or purely generative approaches, there are also promising hybrid architectures possible. These include, as discussed in \cref{sec:terminology:hbyrid_models}, generative models that use discriminatively trained feature embeddings as observation spaces. In such an architecture, physics-based generative models need not be unrealistically detailed in terms of reproducing pixel-level variation in the image, e.g., due to subtleties of lighting or optical material properties, but instead, an approximate generative model that can capture relevant variation with respect to the feature embeddings would suffice. A recent study provided evidence that such a hybrid architecture captures human behavior in the context of the aforementioned domain of the perception of cloth-draped objects, to a better degree than purely discriminative as well as purely generative models \citep{yildirim_3d_2023}.  Hybrid architectures can also be built by taking advantage of recent advances in differentiable graphics systems – by integrating differentiable generative models with differentiable bottom-up feature extraction to enable novel computational hypotheses about perception and its biological implementation, including feedback, lateral, and top-down connections.

\subsubsection{One or few shot generalization} 
\label{sec:empirical:generalization}

\paragraph{Phenomena} 

\textit{One- or a few-shot generalization measured using classification performance}: Humans, including even young children, can learn about new visual categories based only on a few examples of that category – sometimes even a single example of the target category suffices for broad generalization. We review two recent benchmarks that systematically measured human participants’ classification performance in such limited training regimes, which are often referred to as few-shot and one-shot generalization. \citet{lake_human-level_2015} presented the Omniglot dataset, which consists of hand-drawn characters across 10 different world languages, and measured human performance in a 20-way one-shot classification task, which involved finding a letter from a set of 20 others that is of the same alphabet as a target letter. \citet{lee_how_2023} provided a benchmark based on complex, unfamiliar 3D objects, measuring people’s performance of learning these objects under one-shot as well as few-shot settings, across a variety of visual perturbations such as blur, rotation, and pixel deletion. On both benchmarks, people showed remarkable one- and few-shot learning abilities. 

\vspace{3mm}

\textit{One- or a few-shot generalization measured using a drawing task}: Another task, beyond classification, used to evaluate one-shot generalization is a kind of a production task – asking participants to draw new object instances that they believe would belong to the same category as the studied example. \citet{lake_human-level_2015}, in addition to the classification task mentioned above, also asked participants to draw new instances of an alphabet after presenting participants with a few instances of that alphabet. \citet{tiedemann_one-shot_2022} asked participants to freely draw new instances of an object category (defined by its shape) after showing merely a single example. In both cases, humans’ drawing patterns showed abundant variability but also striking commonalities, indicating a rich generative ability to use visual categories learned from even a single example.

\paragraph{Discriminative Interpretation}

What underlies people’s ability to learn about new visual categories so rapidly? One possibility is that such learning simply falls from fast synaptic adaptation over discriminatively trained feature hierarchies. \citet{lee_how_2023} presented such an interpretation, screening several state-of-the-art convolutional neural networks under a number of simple learning rules, leading to a large pool of concrete candidate models. In the few-shot setting (i.e., less than 10 training examples) of the object learning benchmark they proposed, they found humans to be faster learners relative to the strongest CNNs. Moreover, in the one-shot generalization setting (from an unobstructed view of an object to viewing conditions with various kinds of perturbations), they found that humans outperformed all model variants they screened. 

Another discriminative approach is based on the idea of meta-learning, or learning-to-learn, in which an appropriate neural network architecture is put through a training regime to have parameters that can be fine-tuned on new tasks in a data-efficient manner. For example, \citet{finn_model-agnostic_2017} introduced the MAML architecture, which consists of a set of pairs shared across all tasks, and a set of task-specific parameters that can be rapidly fine-tuned on a given new task to be learned. Finn et al. showed strong performance on a number of few-shot learning scenarios.

\paragraph{Generative Interpretation}

Few-shot generalization can be understood as making inferences in a generative model underlying the training examples and using the posterior predictive distribution under this generative model to accomplish broad generalization. The more expressive the generative model, the richer the inferences are expected to be, alongside a need for fewer training items. Thus, it is the extent and validity of the modeling assumptions in a generative model that determines the patterns of one-shot and few-shot generalization. Hierarchical Bayesian models formalize the problem of learning at multiple levels, such that the learning based on a few examples can be transferred or generalized broadly through the higher stages of the hierarchy. \citet{lake_human-level_2015} introduced such a hierarchical Bayesian learner (referred to as “Bayesian Program Learning”) and showed that this model captured people’s one-shot learning abilities, including classification and drawing-based production tasks, in the Omniglot dataset. \citet{ellis_dreamcoder_2023} developed this perspective to a more general, more powerful framework applicable in many relevant domains of visual cognition.

\subsection{Neuroscientific observations}

\subsubsection{Diversity of feature tuning in visual cortex} 

\paragraph{Phenomena} 

The primate visual system includes multiple functionally distinct visual areas that each map a distinct set of features across visual space. The local, detail-sensitive features in primary visual cortex have been extensively characterized, as have the object-related features in more anterior ventral and dorsal visual cortex. Between V1 at the occipital pole and anterior visual areas in the parietal and temporal visual areas are a set of intermediate maps (V2-V4, V3ab, LOC) that are poorly explained by V1-like features and do not show evidence for explicit object selectivity. These visual areas map their diverse features at different levels of spatial resolution, and on average, receptive field sizes tile larger areas of visual space in more anterior regions. For a fixed eccentricity receptive field size \citep{kay_identifying_2008} and spatial frequency preference varies substantially across these brain areas \citep{henriksson_spatial_2008} 

Identifying computational principles that might explain the many different types of representation the visual cortex maintains is a central challenge in visual neuroscience. The discriminative and generative frameworks imply quite different explanations and vary in terms of their ability to learn representations from visual input that resemble those found in primate visual cortex. 

\paragraph{Discriminative interpretation}
Deep discriminative models trained to categorize objects (among other computer vision objectives) learn a hierarchy of representations that accurately predict brain activity in primate visual brain areas. Interestingly, the sequence of layers in many deep discriminative models order primate visual areas along a roughly posterior to anterior axis, so that lower layers best predict activity in V1, deep layers best predict activity in IT, and intermediate layers best predict activity in V2, V3, V4 \citep{yamins_performance-optimized_2014, khaligh-razavi_deep_2014, guclu_deep_2015, cichy_comparison_2016, eickenberg_seeing_2017}. These discriminative networks thus induce a functional hierarchy of brain areas that is roughly consistent with the anatomical hierarchy of Felleman and VanEssen (\citeauthor{felleman_distributed_1991}, \citeyear{felleman_distributed_1991}, but see \citeauthor{st-yves_brain-optimized_2023}, \citeyear{st-yves_brain-optimized_2023}). This finding suggests that the diversity of representations is caused by systematic increases in the computational path length \citep{peters_neural_2022} of visual areas along a functional hierarchy. Diversity of representations happens “along the way” to computing the endpoint representation that optimizes the discriminative objective function, and is necessary to the extent that depth is necessary for learning or computing this endpoint representation.

\paragraph{Generative interpretation}
It has long been known that sparse and shallow generative models of natural scenes learn representations that resemble the Gabor-like receptive fields of neurons in primary visual cortex \citep{olshausen_sparse_1997}. The findings above revealed that discriminative networks can also learn representations that predict and, when visualized, resemble representations in early visual areas too. Furthermore, initial efforts to map representation in neural networks onto brain activity showed that discriminative models tended to yield more accurate predictions of brain activity \citep{khaligh-razavi_deep_2014}. However, as generative models have developed, they have also become repurposed as state-of-the-art predictive encoding models of visual cortex. In particular, diffusion generative models conditioned on semantic descriptors \citep{radford_learning_2021} induce a diverse set of representations that predict activity across visual cortex with high accuracy as well \citep{takagi_improving_2023}.  Thus, there are currently both discriminative and generative models that learn representations that predict brain activity and explain representational diversity. It is also important to note that self-supervised models--which are neither purely discriminative or generative--are emerging as another source of features for predicting brain activity \citep{konkle_self-supervised_2022}.

Interestingly, diffusion models offer a different justification for the diversity of representations than discriminative models. Transformations applied in the $o$ to $z$ are simple additions of unstructured noise to the image that incrementally eradicate first higher than lower spatial frequency information \citep{rissanen_generative_2023}. Taken as a model of vision, the role of the generative transformation (applied in the $z$ to $o$ direction) is to denoise representations that result from a completely random feedforward process--that is, map an image corrupted by noise into a slightly less corrupted image. The mapping can be conditioned upon fixed or “clamped” information about any aspect of the image. Thus, diversity of representation results from the spatial scale of the information that is recovered by the denoising transformation, and the influence of the conditioning information on this transformation.

The fact that both discriminative and generative models as well as models with different training objectives all yield similar cortex-like feature tuning suggests that either (1)  feature tuning across visual cortex reflects the statistics of the visual world and might be shared among many (if not all) successful models of natural visual stimuli, or that (2) extant models for predicting brain activity are not accurate enough to adjudicate between the two frameworks, or are tested on stimulus sets that do provide the sensitivity needed to adjudicate between them \citep{golan_controversial_2020}.

\subsubsection{Visual neural architecture} 
\paragraph{Phenomena} 

Light entering the eyes of a primate is optically focused on the retina where it is transduced into neural signals. It is processed within the retinae and across a sequence of stages including the lateral geniculate nucleus (LGN), primary visual cortex (V1), and higher-level visual cortical areas, including V2 and V4, as well as regions specialized for particular classes of objects, such as faces. These stages form what is known as the ventral stream, which is thought to serve the analysis of the detailed texture and shape of the objects and surfaces the scene is composed of. The complementary dorsal stream is thought of as a somewhat separate set of areas representing global spatial information and contributing to attentional processes. 

\textit{Hierarchical conception.} The primate visual system is often characterized as a hierarchy of regions, a view that emphasizes the sequence of stages and feedforward computation. Consistent with the notion of a hierarchy, visual signals must pass through more synapses and require more time to reach higher cortical areas. Neurons in these areas have larger receptive fields and appear to explicate information about the categories of objects, abstracting from low-level visual features. Connections can be categorized into feedforward (FF, from lower to higher areas), lateral recurrent (LR, within an area), and feedback (FB, from higher to lower areas), not just on the basis of latency of visual responses and the level of abstraction of the neural tuning, but also on the basis of the anatomical microstructure: the projections' cortical layers of origin in the source area and of synaptic termination in the target area. The hierarchical conception emphasizes the placement of each area along a single dimension (its level) while accommodating the notions of feedback and lateral recurrent signaling. Both feedforward and lateral recurrent signaling is already present in the retina. There is no feedback from the LGN to the retinae. However, the LGN receives feedback from cortex. Lateral recurrent and feedback connections are ubiquitous within cortex.

\textit{Interactive conception.} The notion of hierarchy implies that every stage of processing can be placed along a single dimension that determines its synaptic distance from the retina, the level of abstraction of the visual representation, and the classification into FF and FB of its incoming connections. This conception is weakened by a number of anatomical and physiological observations. Anatomically, a given area receives input through a multitude of paths with fewer and more synapses, reflecting abundant FF connectivity that skips hierarchical levels as well as recurrent paths. The microanatomical classification of projections into FF and FB are approximately, but not perfectly consistent with a hierarchy. The characterization of neural tuning does not support an unambiguous and precise placement of regions in a one-dimensional hierarchy, either. Lateral connectivity supports recurrent local processing within areas, and about two-thirds of all possible pairwise connections between areas exist with projections usually present in both directions in similar densities. 

Physiologically, a given area's response latencies vary substantially, forming a wide distribution, and these distributions overlap between areas thought to reside at different levels of the hierarchy. Neural tuning also evolves with the time after stimulus onset, with the same neurons apparently explicating information more abstracted from the image features at larger latencies after stimulus onset.

These anatomical and physiological facts suggest a more interactive conception in which the visual system appears as a tangle of local processors that converge on an interpretation of the sensory evidence by exchanging messages and updating their local representations. In such a conception, FF and FB connections may not be categorically distinct to the same degree in their structure and function.

\paragraph{Discriminative interpretation}

The feedforward aspect of these observations is naturally compatible with discriminative models that often consist of multiple steps of transforming visual inputs on the retina into high-level representations. Some of these models also benefit from operations -- like divisive normalization -- that might be implemented in the brain using lateral connections. Feedback connections from higher to lower visual areas, or bidirectional connectivity in general, cannot be explained by current discriminative models without invoking other functions. For instance, feedback signals may represent learning signals that continually adapt the feedforward computation in the light of visual experience, and feedback. Alternatively, feedback signals might communicate attentional signals, helping to efficiently allocate metabolic costs in terms of increasing firing rates or decreasing variability where important, and decreasing them otherwise.

\paragraph{Generative interpretation}

The observations are compatible with the hypothesis that the brain performs iterative inference in a generative model. Under the assumption that neurons in each area represent posterior beliefs about latents in a generative model, iterative algorithms like MCMC sampling or message passing demand bidirectional communication between ``neighboring'' areas (those representing latents who are in each other's Markov blanket) to communicate information about current beliefs. Such algorithms are also compatible with recurrent connectivity within a cortical area, for instance implementing explaining-away computations \citep{lee_hierarchical_2003}. From this perspective, the output of the retina (or possibly LGN) serves as an observation on which beliefs about latents represented by cortical areas are conditioned. 
Whether visual areas are indeed best thought of as organized hierarchically or as an interactive network without a clear hierarchy is not critical and only sheds light on whether the underlying generative model is hierarchical or not.

\subsubsection{Amodal completion in the visual cortex} 
\paragraph{Phenomena} 
The primate visual system automatically fills in occluded parts of an object. This process is known as \textit{amodal completion} \citep{michotte_nouvelle_1951} and is based on prior knowledge and context. Behavioral evidence from priming studies in humans suggests that amodal completion smoothly extends represented edges behind occluders \citep{kellman_theory_1991} and completes representations of partially occluded shapes symmetrically \cite{buffart_coding_1981}. Such completion is impenetrable by high-level object recognition, as evidenced by completion percepts, which can be explained by low-level symmetry and shape cues but conflict with higher-level knowledge about the world (e.g., object appearances) and context (e.g., the memory of the previously seen, unoccluded object) \citep{kanizsa_amodale_1970}. Activations representing the amodally completed parts have been observed in monkey V1 neurons \citep{sugita_grouping_1999, rauschenberger_temporally_2006}. Amodal completion effects in fMRI have been observed in early visual, extrastriate, and higher-level visual areas \citep{rauschenberger_temporally_2006, weigelt_separate_2007, thielen_neuroimaging_2019, smith_nonstimulated_2010}. 
The neural response of amodally completed representations tends to be slightly delayed compared to the unoccluded condition, suggesting lateral or top-down propagation from the unoccluded to the amodally completed parts \citep[e.g.,][]{lee_dynamics_2001, tang_spatiotemporal_2014, rajaei_beyond_2019}.

Completion effects reflect `world knowledge' that is not present in the current visual input. Such world knowledge can be present in systems with discriminative or generative goals (training). However, the functional significance of the completion effects for inference can be interpreted differently under the discriminative and the generative framework.

\paragraph{Discriminative interpretation}

A discriminatively trained RNN might learn lateral connectivity profiles between edge detectors that reflect the statistics of shapes in the world \citep{geisler_visual_2008} as long as such representations serve the discriminative objective. During inference, lateral information flow between edge detectors of two collinear but disjoint line segments may amodally complete the missing line segment. This process would be a form of denoising serving a functional purpose for downstream computations (e.g., classification) and may also be interpreted as ``emergent local generative inference'' in a discriminative model (see ``hybrid models'' in \cref{sec:terminology:hbyrid_models}). The type of prior information that can be expressed in lateral connection is limited, in line with the fact that amodal completion is limited to simple lines and principles of symmetry.

\paragraph{Generative interpretation}

The existence of amodal completion phenomena in behavior and higher-level cortices is well explained within the generative framework. Higher-level visual areas (e.g., LOC) represent inferences about objects in the input according to the statistics of the world and the current input $p(\mathbf{z}|\mathbf{o})$. Inferences/beliefs at the object level imply beliefs about the occluded parts of an object. In a normative solution, such beliefs would be probabilistic and multi-modal, reflecting the distribution of possible completions compatible with the current input, the temporal and cognitive context, and the world statistics $p(\mathbf{o})$. 

Several perceptual amodal completion phenomena seem inconsistent with the normative ideal and may be informative about the brain's approximation strategies. For example, the amodal percept can sometimes reflect local completion via lower-level cues (e.g., continuation of lines or symmetric completion of shapes), conflicting with the higher-level, global inference about the scene (e.g., global recognition of the occluded object). Rather than achieving inferential consistency between all hierarchical levels, the visual system might employ inferential ``clamping'', by which some lower-level inferences are not conditioned on top-down, higher-level inferences or cognitive context (i.e., cognitively impenetrable, see \cref{sec:empirical:perception_is_impenetrable}).

Neural amodal completion effects in early visual areas might be understood as achieving inferential consistency between local (e.g., V1) and object-level (e.g., LOC) inferences. The receptive field of units in lower visual areas such as V1 may only receive input from the occluding object but still represent beliefs about the occluded object that were conveyed by higher-level areas through top-down connections. These beliefs could be considered mere by-products of inferential consistency in a hierarchical generative model. However, such beliefs might also be functionally relevant for future inferences by anticipating potential future dis-occlusions of the partially occluded object (e.g., via the motion of the object or the observer).

\subsubsection{Spontaneous activity} 
\paragraph{Phenomena} 
Neural activity in visual cortical areas does not cease to exist in the absence of external sensory inputs, e.g., when eyes are closed, or open in absolute darkness. This activity is called spontaneous activity and has been measured using intrinsic imaging \citep{tsodyks_linking_1999}, and extracellular recordings \citep{kenet_spontaneously_2003, berkes_spontaneous_2011}. Importantly, it is structured both spatially (neural variability is correlated between different neurons) and temporally (within the same neuron and between different neurons, across time). Interestingly, this structure resembles the structure of activity that is evoked by external inputs \citep{tsodyks_linking_1999,berkes_spontaneous_2011}. Moreover, the structure inherent in spontaneous activity changes after eye-opening such that after a sufficiently long time after eye-opening the spontaneous activity appears statistically indistinguishable from average evoked activity \citep{berkes_spontaneous_2011}. 

\paragraph{Discriminative interpretation}

There are at least three potential sources of variability in a discriminative model of the brain: (1) input variability due to photo receptor, or other retinal, noise, (2) private neural variability in each unit, e.g., due to stochasticity in the spiking threshold, and (3) synaptic variability. Even in a purely feedforward neural network, the resulting variability in each layer will likely result in structured variability in subsequent layers. Furthermore, the structure of this variability will change as the connection weights change with learning. However, whether such variability indeed resembles average evoked activity as found empirically or whether its match to average evoked activity increases with learning is still an open question.

\paragraph{Generative interpretation}

In models constructed from the generative perspective, posteriors are not just represented at the top level, but also at earlier stages of sensory processing. Under the assumption that the recorded sensory neurons represent posterior beliefs and that these beliefs are represented in neural activity by a linear distributional code \citep{lange_task-induced_2022}, one can show that the neural activity in the absence of sensory inputs (spontaneous activity) should match the evoked neural activity averaged across inputs representative of the natural environment \citep{berkes_spontaneous_2011}.

\subsubsection{V1} 

\paragraph{Phenomena}

In addition to considering individual phenomena that may involve multiple cortical areas as done above, it may be productive to consider multiple phenomena in the context of a single area. Below we will discuss existing models of area V1 from the perspective of discriminative and generative approaches.

\paragraph{Discriminative interpretation}

Much existing work on modeling neural activity in V1 focuses on the relationship between visual inputs and the mean responses of individual neurons. 
It models spike counts statically, as drawn from a Poisson distribution with a mean given by a function $f$ of the stimulus:
\begin{equation*}
    r_i=\mathrm{Poisson}\left[f(\mathbf{o}; \mathbf{W})\right]
\end{equation*}
where current state-of-the-art models use feedforward DNNs to approximate $f$ \citep[BrainScore,][]{schrimpf_integrative_2020}. The parameters $\mathbf{W}$ of these DNNs can be learned either from neural data alone or by using the lower layers in a DNN that has been trained on object categorization \citep{cadena_deep_2019}.


A complementary line of research focuses on the neural dynamics that arise from the fact that cortical neurons within an area are recurrently connected:
\begin{equation*}
    \frac{r_i}{{\rm d}t} = F\left[\mathbf{o}; \mathbf{r},\mathbf{W}\right].
\end{equation*}
Much research has been devoted to understanding the conditions on $F$ and $\mathbf{W}$ that yield dynamics in agreement with known empirical measurements like stimulus-dependent neural variability, and a balance of excitatory and inhibitory inputs to individual neurons \citep[reviewed in][]{ahmadian_what_2021}. 

Insofar as these two classes of models can be interpreted as implementing probabilistic computations, they both can be interpreted as discriminative models since they compute a $p(\mathbf{z|o})$ and contain no obvious generative computations.

\paragraph{Generative interpretation}

A complementary series of studies has modeled V1 activity as reflecting generative inference. \cite{olshausen_sparse_1997, olshausen_emergence_1996} proposed that V1 neurons implement probabilistic inference in a generative model of natural images. The inference algorithm in their model entails recurrent dynamics in which neurons representing latents associated with similar features suppress each other (`explain each other away') -- a prediction for which evidence was recently presented \citep{chettih_single-neuron_2019}. How exactly such probabilistic computations are implemented in neural circuits is still a matter of debate \citep{beck_competing_2020}. Importantly, \cite{olshausen_sparse_1997} and \cite{rao_predictive_1999} proposed implementations based on `predictive coding' while \cite{lee_hierarchical_2003} proposed that neurons in multiple visual cortical areas represent beliefs about latents with inter-cortical signals representing conditional probabilities about the latents in the respective other cortical area. \cite{hoyer_interpreting_2002} and \cite{fiser_statistically_2010} further proposed that neural activity represents probabilistic beliefs in the form of samples, providing a normative explanation for neural variability. In particular, \cite{berkes_spontaneous_2011}, \cite{orban_neural_2016}, and \cite{banyai_stimulus_2019} showed that neural sampling in a generative model could explain many aspects of how neural variability and covariability depend on the nature of the visual input.  \cite{haefner_perceptual_2016} and \cite{lange_task-induced_2022} showed how the assumption that V1 responses represent posterior beliefs in a generative model could also explain the task-dependence of their correlated variability \citep{cohen_context-dependent_2008,bondy_feedback_2018}. Finally, \cite{echeveste_cortical-like_2020} showed that optimizing the parameters of a recurrent neural network containing both excitatory and inhibitory neurons for generating samples from the posterior in a generative model yielded a network that was a special case of the dynamical models described above as compatible with the discriminative framework.

Overall, the fact that the best current models of firing rates can be interpreted as the bottom part of discriminative models of higher-level vision suggests that the phenomena that they miss -- variability, recurrent dynamics \& feedback -- may not be very important for inference, and might instead reflect details of the biological implementation that can be ignored on the computational level. To definitively conclude whether V1 activity reflects discriminative or generative inference requires a firmer understanding of their dependence on top-down signals, and whether V1 activity is compatible with the assumption that it represents beliefs that depend on higher-level beliefs as predicted by the rules of probability.

\newpage

\renewcommand{\itemplacement}{ht}
\input{display-items/box-identifying-the-framework}

\newpage

\newpage

\input{display-items/box-selected-questions}

\newpage

\section{An integrative research program to reveal the algorithms of primate vision}

How can we transcend the dichotomy and understand how the primate brain combines the advantages of discriminative and generative computations? The most enlightening experiments will engage both perspectives, rather than being anchored in just one of them. We should design experiments to adjudicate among a wide range of models that include discriminative, generative, and hybrid models. In this section, we propose ideas for task design, computational modeling, behavioral experimentation, and neural activity measurements that seem to us most likely to drive progress. 

\subsection{Experimental stimuli and tasks}

We noted that the two conceptions of vision occupy diametrically opposed corners of a vast space of potential models. The tasks employed by researchers working under these frameworks similarly appear artificially polarized. Researchers working with discriminative models of perception often use large sets of natural stimuli to train the models and present natural images to their subjects while recording brain activity and behavioral responses. This approach trades control over the stimuli for naturalism and complexity, and relies heavily on training models with large data sets. Researchers working with generative inference models of perception often design simple generative models that serve both to make the stimuli and as components of the perceptual inference models. This approach relies on building the ground-truth generative model into the perceptual inference model and trades off naturalistic complexity for the ability to investigate to what extent perceptual processes approximate the normative ideal of probabilistic inference. Synthetic stimuli, while less natural, can help us probe the ability of primate vision to handle the long tail of rare and novel visual experiences. 

The discriminative perspective often aims to characterize rapid image recognition and categorization, while the generative perspective emphasizes out-of-distribution generalization, learning from a few examples, and multiple tasks (e.g., generation, identifying parts, inferring missing or occluded elements, etc...) \citep{lake_human-level_2015}. Extending the time-scale of visual inference beyond subsecond perceptual trials and incorporating learning into the explanandum may be highly informative for both camps.

Transcending the dichotomy will require experimental stimulus sets and tasks on which discriminative, hybrid, and generative models can compete, both in terms of task performance and as models of primate neural and behavioral dynamics. Computer-implemented tasks \citep[e.g., ][]{bear_physion_2022, gan_threedworld_2021, habitat19iccv} can employ graphics models that can generate a rich variety of complex stimuli, combining the advantages of studies in the two traditions:
\vspace{-4mm}
\begin{enumerate}
    \item \textbf{Larger stimulus sets}: The number and/or dimensionality should go beyond the typically used relatively small sets of hand-selected stimuli and low-dimensional parametric stimulus spaces.
    \item  \textbf{Generativity}: We should have the ability to sample from a continuous distribution of procedurally generated experiences ad infinitum, enabling us to generate novel training and test sets as needed  (e.g., see \Cref{fig:toy-example}a).
    \item \textbf{Naturalism}: The stimuli should be more natural and complex than is typical for studies rooted in the generative framework.
    \item \textbf{World access and control}: We should have control over and access to the factorization of the task world state, which is not usually the case in studies using photographs in the discriminative framework. World access enables explanations of behavior and neural phenomena in terms of task world variables.
    \item \textbf{Scope control}: The ability to sample stimuli with different degrees of typicality (e.g., typical views/images of objects vs. stimuli from the long tail of the distribution that may include uncommon views and pose a generalization challenge to a system).
    \item  \textbf{Out-of-distribution probing}: The ability to introduce distribution shifts over the time-scale of the agent's experience, so as to investigate the ability of models and primates to perform out-of-distribution generalization.
\end{enumerate}

\noindent Our research can continue to be guided by a  normative perspective that starts with the question of how primate vision \textit{should} work. However, this question must be considered in the context of constraints not just on the data (where the normative perspective leads us to probabilistic inference on generative models) but also on computational resources, time, and energy \citep{lieder_griffiths_2020, gershman_computational_2015, russell_provably_1994}. We must consider the complex cost trade-offs that a resource-constrained system faces. Tradable costs include the spatial and temporal complexity of the computation, the energy the computation requires, the amount of data required for recognition and learning, and the costs of perceptual errors. In a task, researchers may want to control not only the overall complexity of the world and the task but also the data available for learning (e.g., by controlling the amount of learning experience) and for inference (e.g., through added noise, occlusions, etc.), as well as constraints on inference and learning (e.g., inference and learning time).

\subsection{Computational modeling}

Computational modeling should explore several promising directions. One important gap is an imbalance in our ability to train deep discriminative and generative models that approximate known cortical architectures. Current discriminative models are easier to train and, therefore, scale more readily to the complexity of primate vision. Better methods for training deep generative models are therefore required. 

The primate visual system is highly recurrent, but the role of recurrence for visual inference is unclear \citep{van_bergen_going_2020}. Both generative and discriminative models of vision can be recurrent, but the role and information conveyed through recurrence are different under both frameworks. Understanding recurrent processing is, therefore, of high relevance for interpreting primate vision in terms of generative and discriminative models. Studying emergent recurrent motifs in task-optimized networks respecting biological constraints is a particularly intriguing research direction because it allows for the discovery of novel, previously unknown, inference algorithms \citep{echeveste_cortical-like_2020,lange_interpolating_2022}.

As researchers, we never have access to the full inference machinery of the primate visual system. It is, therefore, attractive to identify a set of computational motifs that might inform us about the underlying inference framework. Because it is challenging to infer the algorithm of a system by observing limited samples of its activity in isolated experimental paradigms, generative and discriminative models might best be studied in silico first, where the modeler can construct models within a discriminative or generative framework and then try to identify the constructing framework through observation and experimentation on the model. (This amounts to embracing a generative-model-based mode of inference about the algorithm of primate vision at the level of our research.) 

Proponents of either framework often point to normative arguments, reflecting the fundamental bias-variance tradeoff. Generative models can enable more robust inference for noisy and ambiguous inputs than discriminative models. However, the robustness conferred by the (typically misspecified) generative model comes with an inference bias. Discriminative models are more expressive (less biased) and may, therefore, outperform generative models if sufficient training data is available \citep{ng_discriminative_2001}. The human visual system learns its inference model -- either via experience or through evolution -- using much data, but few `labels'. This fact appears to be at tension with a purely discriminative account of primate vision because discriminative models are often trained by supervision using labels. However, this perspective conflates the supervised/unsupervised distinction with the discriminative/generative distinction. With the advent of novel learning objectives (e.g., self-supervised learning), we may see models that can learn from the continuous stream of unlabeled sensory data without acquiring an explicit generative model. A key future direction, therefore, is to explore novel unsupervised and self-supervised learning algorithms.

\subsection{Behavioral and neurophysiological experiments}

Behavioral and neurophysiological experiments should use the same task and stimulus distributions that models are trained and tested with. Several experimental directions stand out as highly informative for the debate at hand. First, experimental paradigms that study ambiguous stimuli or non-stimulus-driven processes.  Techniques that explore multistable phenomena, mental imagery, or spontaneous neural activity, for instance, can help us probe generative computations and distinguish them from iterative, discriminative computations.

At the algorithmic level, visual inference involves an intricate exchange of information between multiple representational stages at a rapid time scale. Candidate models that perform inference using distinct algorithms may settle on similar steady-states and display similar representational statistics across the visual hierarchy. Simultaneous measurements of multiple areas (i.e., at least two) at a fast time-scale may, therefore, be needed to characterize the neural interactions and dynamics that reveal the algorithm underlying primate visual inference.

\section{Conclusion}


Visual neuroscience has recently seen a surge of computational models aimed at unraveling the neural underpinnings of how we perceive our environment. The primary goal of this generative adversarial collaboration was to understand the relationship and bridge the divide between the two main theoretical approaches to biological vision and the corresponding approaches in computer vision. On the one hand, we have discriminative models, which map the visual input directly onto the latent variables to be inferred. On the other hand, generative models represent the joint distribution of observations and latents (often using computations that mimic causal processes in the world) and require complex, typically iterative computations for visual inference.

We delineated the defining characteristics of these two model classes and clarified the terminology, so as to enable a more systematic comparison of different models and their implications. However, we also revealed a considerable gray area of models that do not fit neatly into either the discriminative or the generative category. An intriguing possibility is that the primate brain's visual algorithm is neither purely discriminative nor purely generative. Primate vision may employ a hybrid algorithm that combines the advantages of both paradigms.

We evaluated empirical evidence commonly cited in support of each model class, concluding that many of the empirical results can plausibly be interpreted through the lens of either the discriminative or the generative framework. This further underscores the need for an integrative research program that transcends the dichotomy and embraces the space between the two approaches. Addressing the many remaining open questions (\hyperlink{link-box:questions}{Box 5}) will require experimental tasks and stimuli that combine naturalism and control, can support the development of both discriminative and generative models, and can be used in behavioral and neurophysiological experiments with human and nonhuman primates.


\newpage

\bibliography{references}

\section*{contributions}

All authors contributed to the overall discussion and gave feedback on drafts. Initial drafts: Introduction (NK, BP), Terminology and Scope (BP), Discriminative Perspective (LI, JD, TK), Generative Perspective (RH, TN), Hybrid perspective (NK), Recognition reaction times (LI, RH, NK), Visual imagery and aphantasia (TN), Perception is impenetrable by cognition (LI), Vision generalizes (IY, JT), One or few shot generalization (IY, JT), Diversity of feature tuning in visual cortex (TN, LI), Visual neural architecture (NK, RH), Amodal completion in the visual cortex (BP), Spontaneous activity (RH), V1 (RH), Integrative research program to reveal the algorithms of primate vision (BP), Outlook (RH).

\section*{funding}

B.P. has received funding from the EU Horizon 2020 research and innovation programme under the Marie Skłodowska-Curie grant agreement no. 841578. This work was also supported by the National Science Foundation under Grants No. 1948004 to N.K., and CAREER/IIS-2143440 to R.M.H., by the National Institute of Mental Health under Grant No. R01MH132826 to L.I., the National Eye Institute under Grants No. R01EY023384 to T.N., and R01EY028811 to R.M.H. This work was moreover partially funded by the Office of Naval Research (N00014-20-1-2589, to J.J.D.); (MURI, N00014-21-1-2801, to J.J.D.), the National Science Foundation (2124136, to J.J.D.); (NSFSTC CCF-1231216, to J.J.D.), the Simons Foundation (542965, to J.J.D.); (NC-GB-CULM-00002986-04, to J.J.D.), and the Semiconductor Research Corporation (SRC) and DARPA.









\section*{acknowledgements}

We would like to thank Hossein Adeli, Veronica Bossio, Eivinas Butkus, Wenxuan Guo, Paul Linton, Savannah Smith, Patrick Stinson, and JohnMark Taylor for their helpful comments.

\section*{conflict of interest}

The authors declare no conflict of interest.

\graphicalabstract{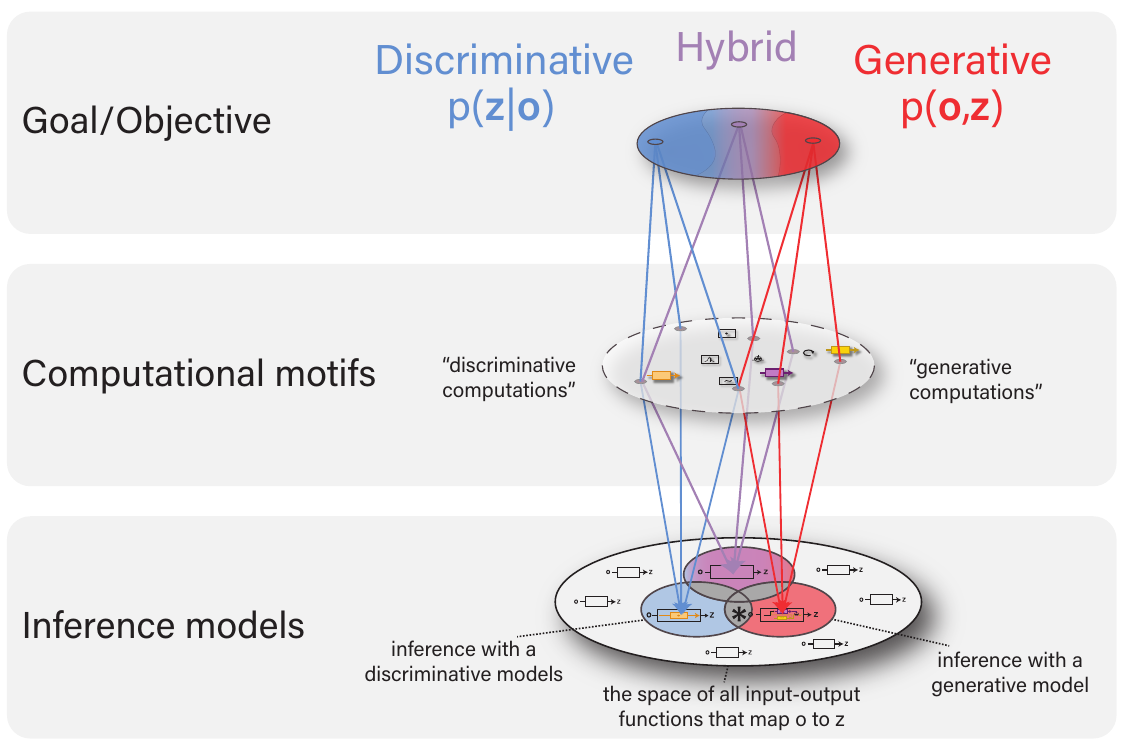}{}

\end{document}

%% file: display-items/box-models.tex
\begin{abox}{p}{BOX: The different types of `model'}
\begin{wrapfigure}{l}{0.4\textwidth}
\includegraphics[width=.4\textwidth]{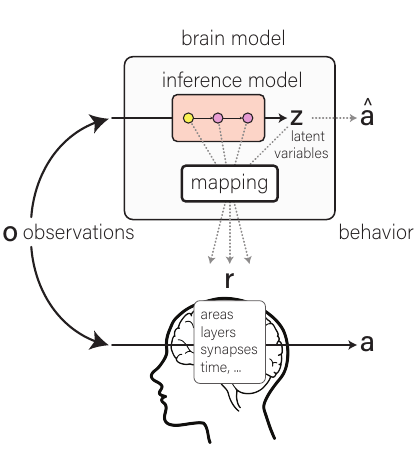}
\end{wrapfigure}
\hypertarget{link-box:models}{} \label{box:models}

{\small
\begin{spacing}{0.85}
 A \textbf{\textit{brain model}} is a researcher's model of how observations $\mathbf{o}$ relate to brain responses $\mathbf{r}$ and behavioral responses $\mathbf{a}$. A particular form of brain models are representational models that construe understanding of brain computations in terms of \textbf{\textit{latent variables}}. Latent variable brain models of visual inference consist of two parts: an \textbf{\textit{inference model}} ($o \rightarrow z$) and a mapping model ($z \rightarrow r$). An \textbf{\textit{inference model}} is a model that computes an estimate of the latent variables $\mathbf{z}$ (point or probabilistic estimate) for a given observation $\mathbf{o}$. 
 A \textbf{\textit{mapping model}} relates latent variables to brain responses (individual neurons, activity in a region of interest).
 It is important to only consider simple mapping models and/or a-priori agree on acceptable mapping model classes. The minimal case of a mapping model is a correspondence function that ``assigns'' individual latent variable $\mathbf{z}_i$ to responses of individual units of brain responses $r_j$. Typically, we might allow for some invariance against brain idiosyncrasies by fitting simple (e.g., linear or low complexity) mappings between $\mathbf{z}$ and brain responses \protect{\citep{ivanova_beyond_2022}}. We distinguish between a \textbf{\textit{brain model}} (a researcher's model of the brain) and a \textbf{\textit{world model}} (a ``generative model'' that captures a prior over possible scenes and/or causal mechanisms in the world that give rise to the retinal data). 
 A brain model might comprise a world model as part of its inference model.
\end{spacing}
}
\end{abox}

%% file: display-items/fig-toy-example.tex
\begin{figure}[p]
\hypertarget{link:fig:toy-example}{}
\begin{center}
\includegraphics[width=\textwidth]{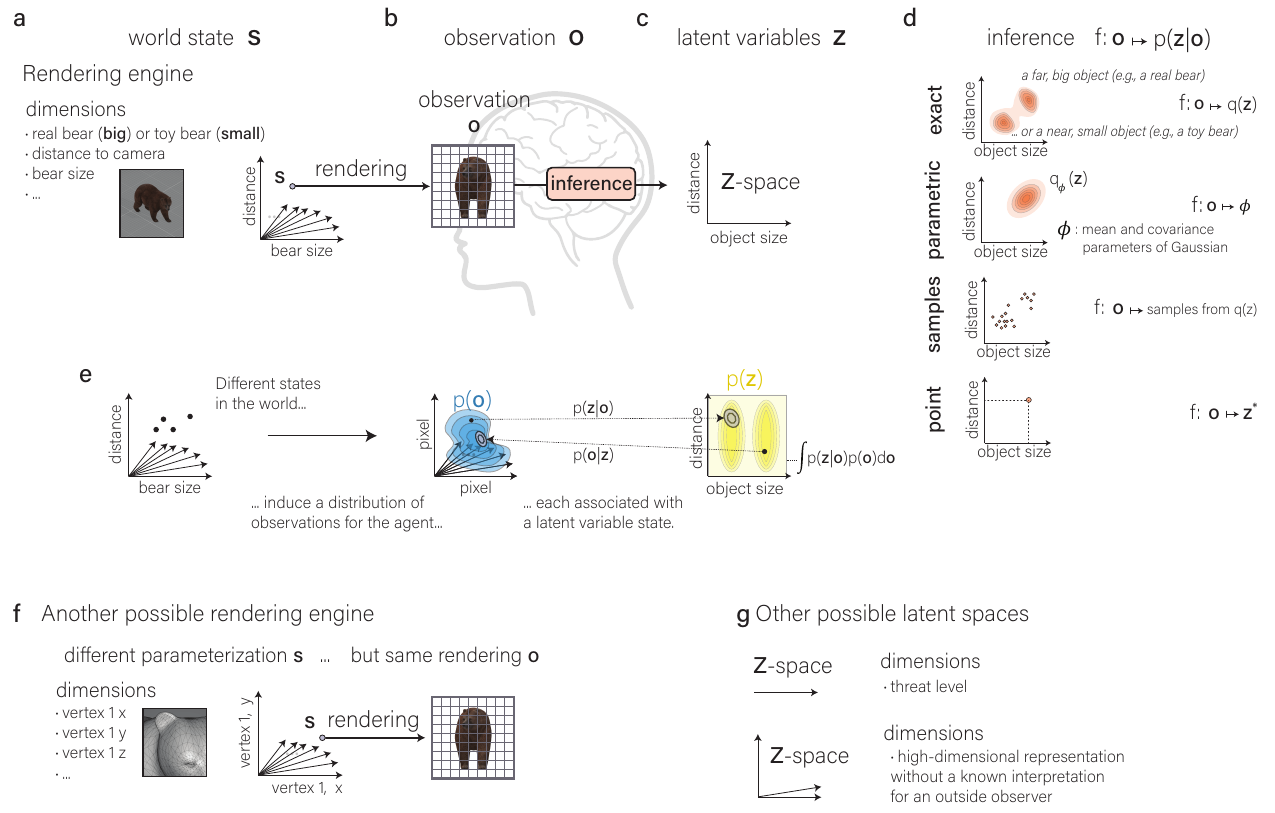}
\end{center}
\caption{
\textbf{Toy example of the visual inference problem.} (\textbf{a}) As an example, we'll consider a world we can fully control that generates the sensory experiences for an agent. Our example world consists of a simplified environment for which a graphics engine renders a single object (a bear). The graphics engine has only a few parameters: whether the bear is a real bear or a teddy bear, a size distribution conditioned on the type of bear (big for real, small for toy), and the distance of the bear to the camera (i.e, the observer). 
(\textbf{b}) From the world state $\mathbf{s}$, the graphics engine renders a sensory pattern on the retina of the agent: the observation $\mathbf{o}$. 
(\textbf{c}) The agent infers some latent variables $\mathbf{z}$ that reflect aspects of the world, here the size and distance of the object, which can be represented by a two-dimensional vector $\mathbf{z}$ in the agent's mind. 
(\textbf{d}) When presented with a specific input $\mathbf{o}$, the agent infers the corresponding latent variables $\mathbf{z}$. Inference can therefore be understood as a function $f$ which maps each observation $\mathbf{o}$ onto a belief about the latent variables. The belief about $\mathbf{z}$ might contain uncertainty, requiring a probabilistic representation $p(\mathbf{z}|\mathbf{o})$. The agent might compute and represent the exact posterior $p(\mathbf{z}|\mathbf{o})$, a parametric representation $\phi$ (e.g., means and covariances of a multivariate Gaussian), represent the distribution in the form of samples $\mathbf{z} \sim p(\cdot|\mathbf{o})$, or in form of a single point estimate (e.g., the vector $\mathbf{z}^*$ that maximizes $p(\mathbf{z}|\mathbf{o})$. 
(\textbf{e}) The statistical structure of the inference problem: States in the world $\mathbf{s}$ induce a distribution over observations $p(\mathbf{o})$ that an agent encounters. Each observation is associated with a particular latent variable state, giving rise to a distribution over latent variables $p(\mathbf{z}) = \int p(\mathbf{z}|\mathbf{o}) p(\mathbf{o}) d \mathbf{o}$ that an agent can expect to experience in the world. 
(\textbf{f}) Latent variables in the brain need not correspond to the world state. Another rendering engine with a different parameterization of the world (e.g., in terms of the position of individual vertices in space) could generate the exact same observations (and the exact same $p(\mathbf{o})$). This highlights the elusive relationship between $\mathbf{s}$ and $\mathbf{z}$. In particular, it is not warranted to simply assume $\mathbf{z}$ from a particular, hypothesized `world representation'. Instead, it is an empirical question of how the brain's latent variables relate to a researcher's model of the world $\mathbf{s}$. The variables $\mathbf{s}$ in the researcher's mental model (e.g., bear category, bear size) might not correspond to latent variables in the brain's visual inference model (see (\textbf{g}) for other examples of possible latent variables). This raises the fundamental question of how the latent variables $\mathbf{z}$ arise from the agent's interactions with the world. 
} 
\label{fig:toy-example}
\end{figure}

%% file: display-items/box-primer.tex
\begin{bbox}{p}{BOX: A brief primer on inference with discriminative and generative models}
\label{box:primer}
\hypertarget{link-box:primer}{}
\subsection*{Discriminative models, generative models, and discriminative inference}
\begin{center}
\includegraphics[width=1.\textwidth]{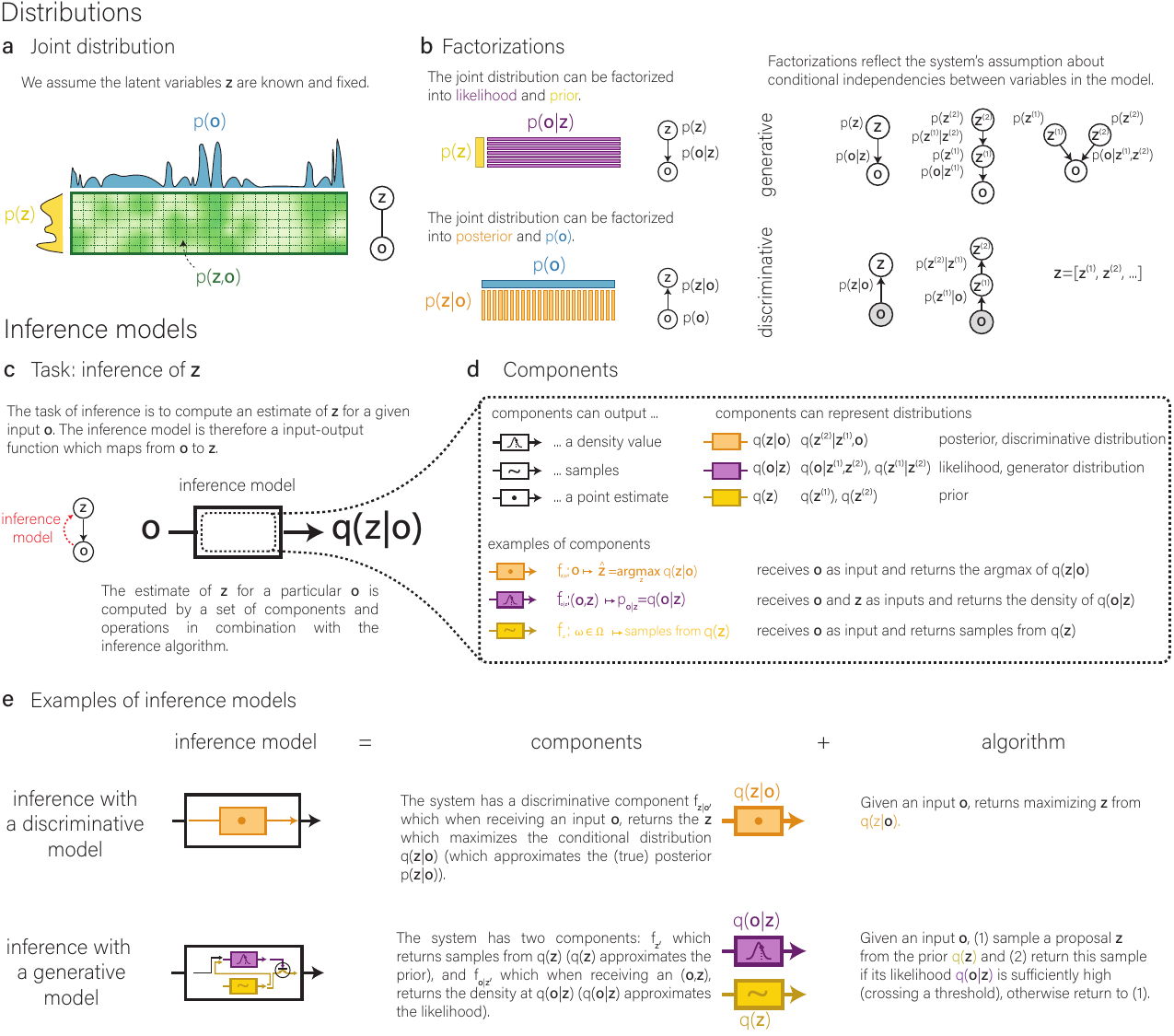}
\end{center}
\vspace{-.7em}
{
\small 
\begin{spacing}{0.75}
(\textbf{a}) A system or agent experiences observations $\mathbf{o}$. We assume that the latent variables $\mathbf{z}$ are fixed (i.e., there is a stable relationship between $\mathbf{z}$ and $\mathbf{o}$). Together with the distribution over observations in the world that the agent encounters $p(\mathbf{o})$, we obtain a joint distribution over observations and latent variables $p(\mathbf{o},\mathbf{z})$ (high-dimensional $\mathbf{o}$ and $\mathbf{z}$ flattened for visualization). (\textbf{b}) The joint distribution can be factorized into $p(\mathbf{o}|\mathbf{z})p(\mathbf{z})$ or $p(\mathbf{z}|\mathbf{o})p(\mathbf{o})$. A system might have a statistical model of the world, i.e. modeling the joint distribution of observations and latent variables. Such a model is a generative model in the statistical sense: it assigns probabilities to observations. Right top: The system might have assumptions about conditional independencies between its variables, which are often expressed in the form of graphical models (Bayes nets). Right bottom: A discriminative model has no model of the joint distribution but might still have conditional independence assumptions among its variables (e.g., conditional-random-field or maximum-entropy models, \protect{\cite{sutton_introduction_2012}}). (\textbf{c}) An inference model is a function that takes an observation $\mathbf{o}$ as input and returns an estimate over the latent variable $\mathbf{z}$, $q(\mathbf{z}|\mathbf{o})$. In the machine learning literature, the inference algorithm is sometimes depicted together with the statistical model as a dotted line (left, red dotted line). Here, we visualize the inference algorithm as an input-output function (black box) with internal components (right, \textbf{d}). These internal components are input-output functions themselves, the can represent distributions (color-coded), and they might have different input-output types (e.g., taking a vector as input and returning a density value or a sample as output). Note, that $q(\cdot)$ refers to model components that implement distributions $p(\cdot)$ (potentially in an approximate way). (\textbf{e}) Top: the objective of the statistical model (representing $p(\mathbf{z}|\mathbf{o})$) aligns with the objective of the inference model (returning an estimate for $\mathbf{z}$ for a given input $\mathbf{o}$). Bottom: In the case of generative models, the objective of the statistical model (i.e., representing the joint distribution over $\mathbf{o}$ and $\mathbf{z}$) does not align with the inference objective. The model therefore needs to be ``inverted'' in order to perform inference with the components (here: likelihood and prior) of the generative model.
\end{spacing}
}

\end{bbox}

%% file: display-items/fig-constructing-inference-models.tex
\begin{figure}[p]
\hypertarget{link-figure:constructing-inference-models}{}
\begin{center}
\includegraphics[width=\textwidth]{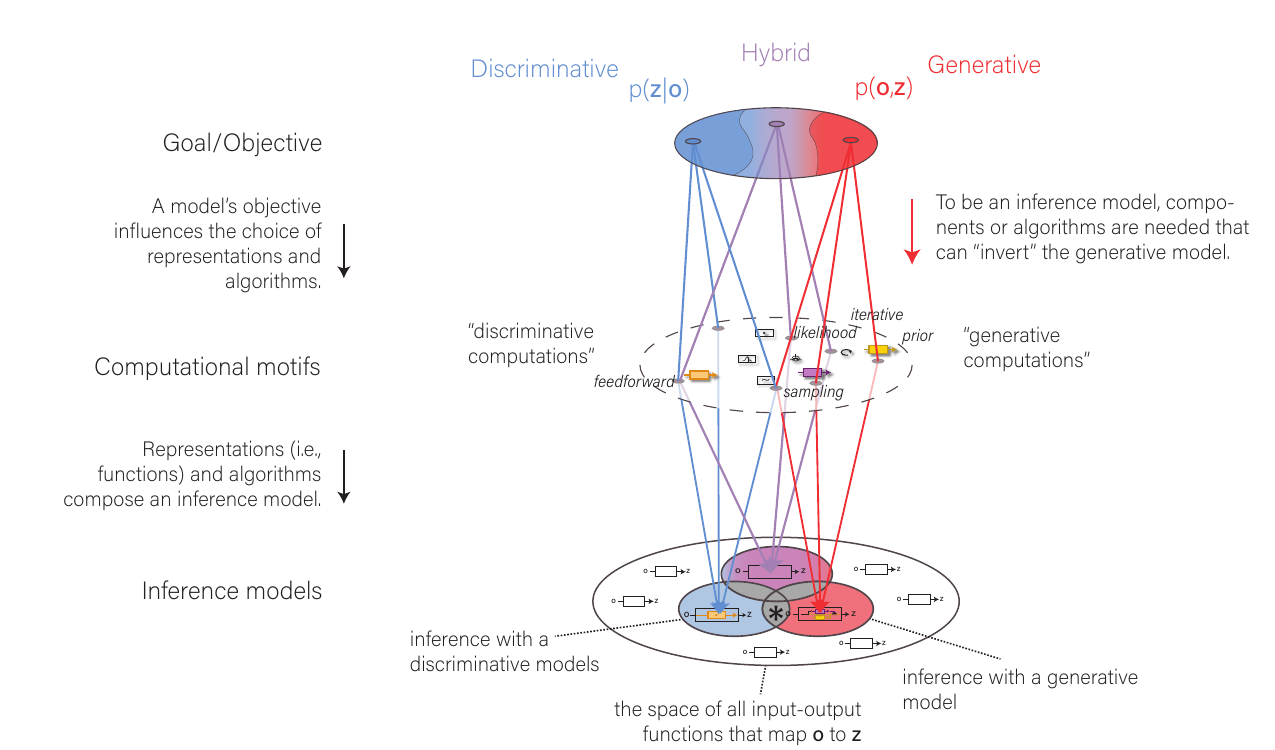}
\end{center}
\vspace{-5mm}
\caption{\textbf{Frameworks for constructing inference models.} We are interested in inference models, i.e. models that map $\mathbf{o}$ onto an estimate of $\mathbf{z}$. ``Discriminative'' and ``generative'' can be seen as frameworks to arrive at hypothetical models of how the visual system performs inference (bottom ellipse: the space of all possible inference models). Model construction starts with defining the system's overall objective (upper ellipse): discriminative ($p(\mathbf{z}|\mathbf{o})$), generative ($p(\mathbf{o}, \mathbf{z})$) or hybrids between generative and discriminative objectives. For example, the generative framework starts by positing that the overall system's objective is to represent the joint distribution over observations $\mathbf{o}$ and latent variables $\mathbf{z}$. Middle ellipse: The choice of goal/objective leads to particular choices for representations and algorithms that implement inference. Inference with a generative model (in contrast to a discriminative model) needs to ``invert'' the generative model such that its components compute an estimate of $p(\mathbf{z}|\mathbf{o})$ (red arrow between upper two ellipses). This difference between the discriminative and generative models biases the choice of algorithms and representations under both frameworks (reflected in the notion of ``generative computations'' and ``discriminative computations''). Importantly, however, both frameworks may lead to overlapping choices of \textit{individual} representational and algorithmic motifs (e.g., discriminative models may be iterative and involve sampling). Lower ellipse: The resulting classes of inference model are distinct by virtue of their construction framework. Possibly, different frameworks may lead to the same inference models (\textbf{*} in the intersection of generative and discriminative inference models). For example, a discriminatively trained RNN might - in principle - learn representations and algorithms that implement inference with a generative model).
}
\label{fig:constructing-inference-models}
\end{figure}

%% file: display-items/fig-examples-of-system-interpretation.tex
\begin{figure}[p]
\hypertarget{link-figure:examples-of-system-interpretation}{}
\begin{center}
\includegraphics[width=\textwidth]{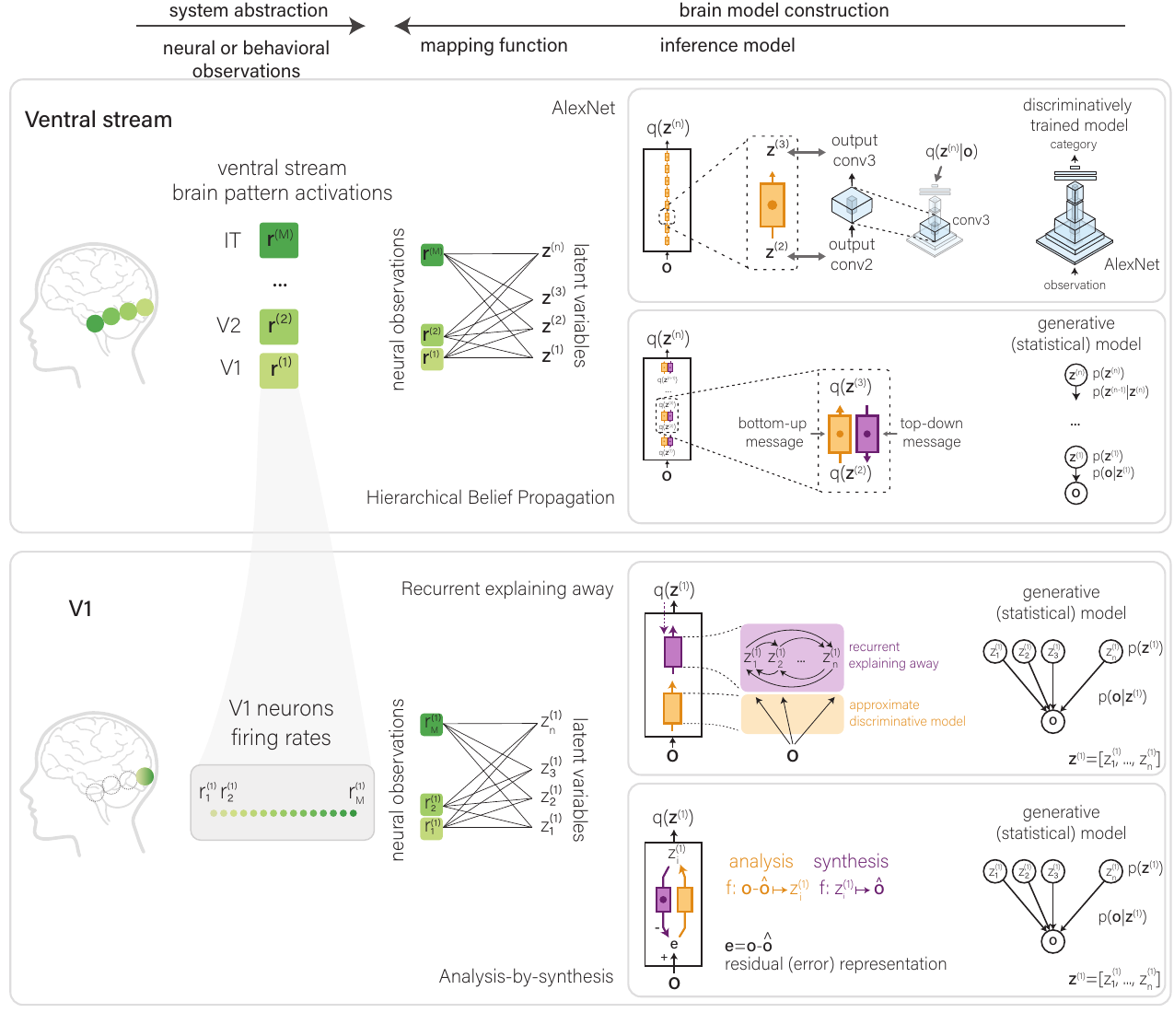}
\end{center}
\caption{\textbf{Examples of interpreting systems in terms of discriminative and generative models.} A brain-behavioral model is an interpretation of a particular system abstraction (e.g., firing rates of V1 neurons, or ventral visual stream pattern activations) that involves an inference model and a mapping function that maps components of the inference model $\mathbf{z}$ to measured neural activity and/or behavior $\mathbf{r}$. Ventral stream (top box): two examples of interpreting brain pattern activations in the primate ventral stream. Top: Individual layers of a trained AlexNet can be interpreted as discriminative components of an inference models. Bottom: Alternatively, these neural representations can be interpreted as corresponding to beliefs in a hierarchical belief propagation network, a model of generative inference suggested by e.g., \protect{\cite{lee_hierarchical_2003}}. Bottom: two examples of interpreting V1 neuronal firing rates. Top: firing rates in V1 can be mapped onto the latent variables of a model, in which inhibitory lateral connections between representational units $z_1, \dots, z_n$ perform a version of explaining away \protect{\citep[e.g., ][]{olshausen_emergence_1996}}. Bottom: an example of V1 neural activations explained via an analysis-by-synthesis model \protect{\citep[e.g., ][]{olshausen_sparse_1997}} that 'inverts' the generative model. Note, that this model includes error units, representing the momentary difference between predicted and observed input, a prediction which can be included into the set of latent variables which are mapped onto neural observations.
}
\label{fig:examples-of-system-interpretation}
\end{figure}

%% file: display-items/fig-spectrum-of-hybrid-models.tex
\begin{figure}
\hypertarget{link-figure:spectrum-of-hybrid-models}{}
\begin{center}
\includegraphics[width=\textwidth]{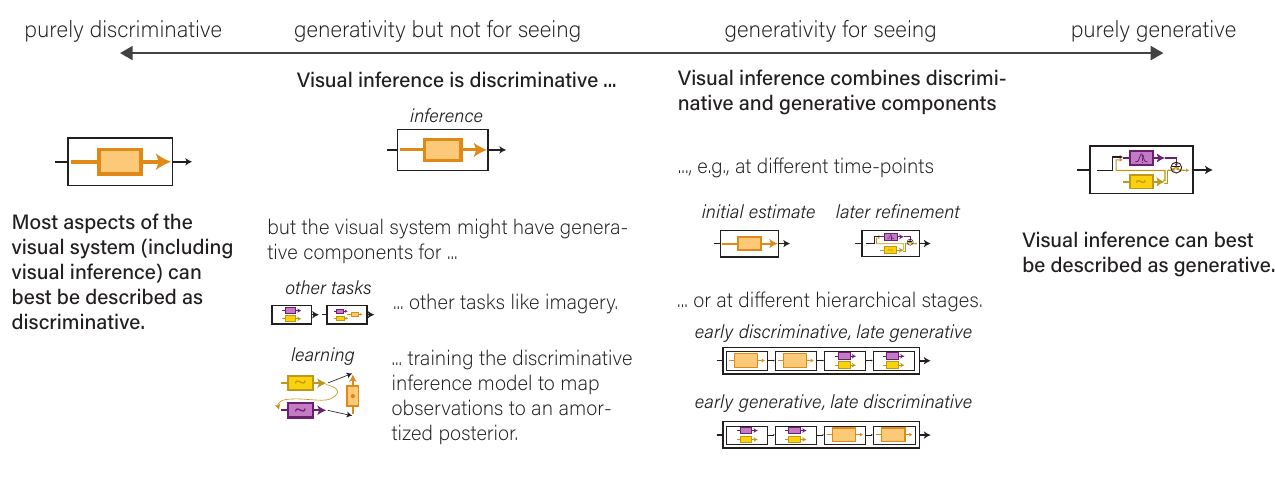}
\end{center}
\caption{\textbf{Spectrum of perspectives on vision.} There is a spectrum of models whose extreme poles are purely discriminative and purely generative models of visual inference (i.e., ``seeing''). A predominant discriminative perspective on visual inference may be compatible with the notion that generativity is involved in other visual tasks or in training the inference component of the visual system (2\textsuperscript{nd} column). The hybrid perspective suggests that the visual system combines discriminative and generative components, which may be identified in time and space (3\textsuperscript{rd} column). 
}
\label{fig:spectrum-of-hybrid-models}
\end{figure}

%% file: display-items/box-imporant-nuances.tex
\begin{abox}{p}{BOX: Important nuances} \hypertarget{link-box:nuances}{} \label{box:nuances}

{\small
\begin{spacing}{0.95}

{\normalsize \textbf{Do generative models generate or reconstruct observations?}}  


Not necessarily. A purely discriminative model (where both computational goal and the probabilistic factorization are discriminative) does not contain the machinery to generate new observations with $\mathbf{o} \sim p(\mathbf{o})$. Moreover, its latent representation may lack the full information needed to reconstruct an image. Models that are generative at the level of the information (i.e., modeling the joint distribution of latents and observations) contain all the information such that one \textit{could} generate new observations or reconstruct observations (i.e., generating observations that are compatible with the currently inferred latent variables). However, turning this model into a generator that samples new observations $\mathbf{o}$ might involve additional mechanisms which are not part of the original model (e.g. if the original model was a model of the density $p(\mathbf{o}, \mathbf{z})$). Hence, a generative model might not involve any representations and computations that appear like generating or reconstructing stimuli.

{\normalsize \textbf{Do discriminative models discriminate (i.e. classify) observations?} }  


Not necessarily. Discriminative models can perform classification (i.e., categorical variable $\mathbf{z}$) or regression (i.e., continuous variable $\mathbf{z}$). In the context of this GAC, we consider $\mathbf{z}$ to be a high-dimensional variable possibly combining categorical and continuous information. So the term discriminative inference does not imply a restriction to classification models but denotes that the goal is to infer $p(\mathbf{z}|\mathbf{o})$.

{\normalsize \textbf{Are discriminative models implemented by feedforward models and generative models implemented by recurrent models?}}


No. Although generic algorithms for inference on generative models often involve recurrence, we can, in principle, perform discriminative inference in the context of a generative model using only feedforward computations (e.g., exactly inverting a linear generative model or approximately inverting any generative model). Conversely, discriminative models can also employ recurrence (e.g., an RNN as a discriminative classifier). Inverting a generative model is often hard or even intractable. Approximate inversion schemes, therefore, often use iterative algorithms that lend themselves to implementation using recurrent mechanisms.

{\normalsize \textbf{Do generative models have richer intermediate representations than discriminative models?}}


In the context of this GAC, we refer to the high-dimensional latent variable $\mathbf{z}$ (e.g., related to activity patterns in IT cortex via a mapping function) as the target of inference and we ask whether inference (of the same target $\mathbf{z}$) is performed with a discriminative or generative model. Hence, $\mathbf{z}$ and the conditional distribution $p(\mathbf{z}|\mathbf{o})$ captured by a model may be identical under both frameworks. However, intermediate representations (e.g., layers in an ANN) in the inference model, which computes $p(\mathbf{z}|\mathbf{o})$ from the input $\mathbf{o}$ may be different between the two approaches. The generative, but not the discriminative, approach computes $p(\mathbf{z}|\mathbf{o})$ using representations that capture the statistical structure of the data (i.e., the joint distribution $p(\mathbf{o}, \mathbf{z})$, e.g., factorized as likelihood and prior). Intermediate representations may, therefore, be richer under the generative framework.



{\normalsize \textbf{Are discriminative models trained with supervision and generative models trained without?}}

Discriminative models can only be trained with supervision (by optimizing parameters $\theta$ to maximize $p_\theta(\mathbf{z}|\mathbf{o})$ for data pairs of observations $\mathbf{o}$ and variables $\mathbf{z}$). Generative models can be trained without supervision (e.g., by maximizing the marginal likelihood $p(\mathbf{o}) = \int_\mathbf{z} p_\theta(\mathbf{o}|\mathbf{z})p_\theta(\mathbf{z})$ for observations $\mathbf{o}$). However, generative models can also be trained with supervision by maximizing the joint likelihood $p(\mathbf{o},\mathbf{z}) = p_\theta(\mathbf{o}|\mathbf{z})p_\theta(\mathbf{z})$ between observations $\mathbf{o}$ and variables $\mathbf{z}$.

\end{spacing}
}
\end{abox}

%% file: display-items/box-identifying-the-framework.tex
\begin{bbox}{p}{BOX: Challenges in identifying the framework} \hypertarget{link-box:identifying-the-framework}{} 
\label{box:identifying-the-framework}
\begin{center}
\includegraphics[width=1.\textwidth]{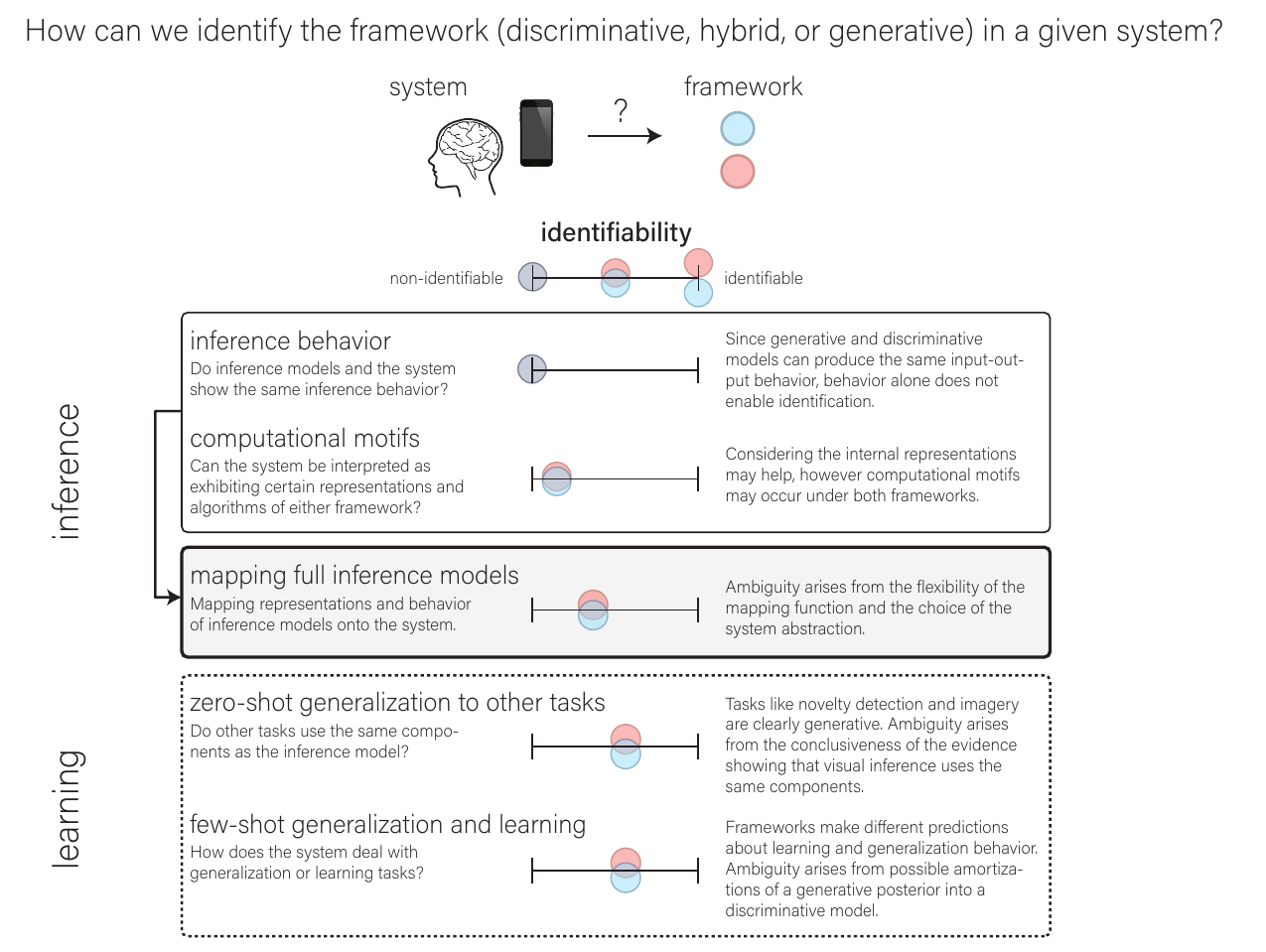}
\end{center}
\vspace{-5mm}
{\small
\begin{spacing}{0.85}
\textbf{What observations or experiments can reveal the framework under which a system was constructed?} Assuming we had access to all components and activations of a system and we could perform all possible experiments, how can we identify the framework under which the inference system (e.g., the visual system) was constructed? Identifying the framework or interpreting the inference of a system (e.g., the visual system or a white box system such as an iPhone) in terms of discriminative, hybrid, or generative frameworks is most promising when mapping the internal representations, algorithmic signatures, and behavior between inference model and system (``mapping full inference models''). Behavior alone or particular computational motifs may not enable us to identify whether a system performs inference within a generative or discriminative framework. The competing frameworks make distinct predictions for generalization behavior (from zero-shot generalization through few-shot generalization to learning new tasks by an existing model). Using other tasks (e.g., novelty detection or imagery) beyond visual inference can inform visual-inference models if we can show that the same components are used for these tasks and visual inference. 
\end{spacing}
}
\end{bbox}

%% file: display-items/box-selected-questions.tex
\begin{abox}{p}{BOX: Selected questions and challenges} \hypertarget{link-box:questions}{} \label{box:questions}

{\small
\begin{spacing}{0.95}


    

    \begin{itemize}

        \item Are there uniquely identifying components of the discriminative versus generative framework? 
        
        \item What is the difference between discriminative inference using recurrence and iterative inference in a generative model? Is this distinction meaningful theoretically? If yes, what would be empirical tests to distinguish them? 
        
        \item Can the space between the discriminative and generative frameworks be understood as a trade-off between domain-specific efficient inference and more generalizable generic inference? 

        \item Do generative models have intermediate representations that are ``richer'' in some sense than discriminative models? 

        \item Is there an inherent indeterminacy, where the computational compromises an inference algorithm makes can always be attributed to either incorrect assumptions about the world that render inference easier or approximations of the algorithm? 

        \item Will alternatives to the existing frameworks be hybrids of both approaches or something entirely novel? Concrete instantiations of hybrid models are needed to compare them quantitatively to both discriminative and generative models of sensory processing.

        \item 
        We motivated both discriminative and generative perspective in probabilistic terms, i.e. in terms of the distributions over latents that the respective models compute. 
        There are at least two unresolved questions in relating probabilistic models to neural activity: First, how do probabilities relate to neural activity \protect{\citep{beck_competing_2020}}? Second, what is the right mapping model when the model of inference is probabilistic?
    \end{itemize}

\end{spacing}
}

\vspace{-.25cm}

\end{abox}